\documentclass[preprint,journal]{vgtc}       
\ieeedoi{10.1109/TVCG.2019.2934810}

\ifpdf
  \pdfoutput=1\relax                   
  \usepackage{graphicx}                
  \DeclareGraphicsExtensions{.pdf,.png,.jpg,.jpeg} 
\else
  \usepackage{graphicx}                
  \DeclareGraphicsExtensions{.eps}     
\fi%

\graphicspath{{figures/}{pictures/}{images/}{./}} 

\usepackage{microtype}                 
\PassOptionsToPackage{warn}{textcomp}  
\usepackage{textcomp}                  
\usepackage{mathptmx}                  
\usepackage{times}                     
\usepackage{cite}                      
\usepackage{tabu}                      
\usepackage{booktabs}                  
\usepackage{multirow}
\usepackage{chngpage}
\usepackage{wrapfig}
\usepackage{color, soul}
\usepackage{diagbox}
\usepackage{dblfloatfix}    
\usepackage{float}
\usepackage{setspace}
\usepackage{enumitem}
\usepackage{amsmath}
\usepackage{amsfonts}
\usepackage{siunitx}
\usepackage[skip=0.2\baselineskip]{caption}
\usepackage[linesnumbered,lined,commentsnumbered]{algorithm2e}
\setitemize{noitemsep,topsep=1pt,parsep=0pt,partopsep=0pt}
\newcommand{\cmo}{\textcolor[rgb]{0, 0, 0}}

\onlineid{1110}

\vgtccategory{Research}
\vgtcpapertype{algorithm/technique}

\title{Towards Automated Infographic Design: \\ Deep Learning-based Auto-Extraction of Extensible Timeline}

\author{
  Chen Zhu-Tian, 
  Yun Wang,
  Qianwen Wang, 
  Yong Wang, 
  and Huamin Qu 
  }
\authorfooter{
\item
C. Zhu-Tian, Q. Wang, Yong Wang and H. Qu are with Hong Kong University of Science and Technology. E-mail: \{zhutian.chen, qwangbb, ywangct\}@connect.ust.hk, huamin@cse.ust.hk.
\item
Y. Wang is with Microsoft Research. E-mail: wangyun@microsoft.com.
}

\abstract{Designers need to consider not only perceptual effectiveness but also visual styles when creating an infographic.
This process can be difficult and time consuming for professional designers,
not to mention non-expert users,
leading to the demand for \emph{automated infographics design}.
As a first step,
we focus on timeline infographics,
which have been widely used for centuries.
We contribute an end-to-end approach that automatically extracts 
an extensible timeline template from a bitmap image.
Our approach adopts a deconstruction and reconstruction paradigm.
At the deconstruction stage, we propose a multi-task deep neural network that simultaneously parses two kinds of information from a bitmap timeline: 
1) the global information, 
\cmo{\emph{i.e.},}
the \emph{representation}, \emph{scale}, \emph{layout}, and \emph{orientation} of the timeline, 
and 2) the local information, 
\cmo{\emph{i.e.},}
the \emph{location}, \emph{category}, and \emph{pixels} of each visual element on the timeline.
At the reconstruction stage, 
we propose a pipeline with three techniques, \emph{i.e.}, \emph{Non-Maximum Merging}, \emph{Redundancy Recover}, and \emph{DL GrabCut},
to extract an extensible template from the infographic, 
by utilizing the deconstruction results. 
To evaluate the effectiveness of our approach, 
we synthesize a timeline dataset (4296 images) and collect a real-world timeline dataset 
(393 images) from the Internet.
We first report quantitative evaluation results of our approach over the two datasets. 
Then, we present examples of automatically extracted templates
and timelines automatically generated based on these templates to qualitatively demonstrate the performance. 
The results confirm that our approach can effectively extract extensible templates from real-world timeline infographics.
} 


\CCScatlist{ 
 \CCScat{K.6.1}{Management of Computing and Information Systems}%
{Project and People Management}{Life Cycle};
 \CCScat{K.7.m}{The Computing Profession}{Miscellaneous}{Ethics}
}

\teaser{
  \centering
  \includegraphics[width=\linewidth]{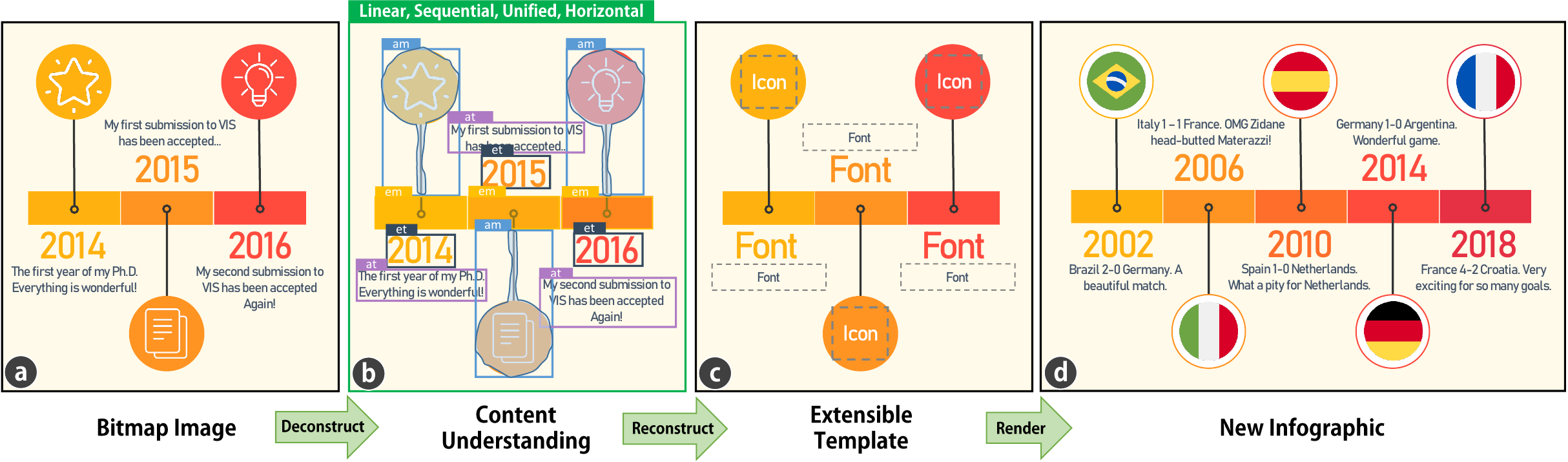}
  \caption{An automated approach to extract an extensible timeline template from a bitmap image.
  a) Original bitmap image;
  b) Content understanding including global and local information of the timeline;
  c) Extensible template contains editable elements and their semantic roles; 
  d) New timeline \cmo{(with mock-up colors)} automatically generated with updated data.
  }
    \label{fig:teaser}
}

\vgtcinsertpkg

\begin{document}


\firstsection{Introduction}
\maketitle
Graphic designers have been producing infographics in a variety of fields, 
such as advertisement, business presentation, and journalism, 
\cmo{because of their effectiveness 
in spreading information~\cite{Harrison2015, Wang2018c}.}
To inform data context and engage audiences,
infographics are often embellished with icons, shapes, and images in various styles~\cite{Kim2017}.
However, creating infographics is demanding.
Designers should 
consider
not only perceptual effectiveness but also aesthetics, memorability, and engagement~\cite{ Borkin2016, Haroz2015}. 
Researchers have introduced design tools~\cite{Kim2017, Xia2018, Wang2018c}
to alleviate the burden of infographics creation
\cmo{by automating some processes (\emph{e.g.}, visual encoding).
However,
these tools 
require users to manually initialize
most of the design (\emph{e.g.}, drawing graphical elements).}
The process remains difficult and time-consuming, 
especially for laymen,
leading to the \cmo{demand} for \emph{automated infographic design}.

Using \cmo{templates} is an effective approach to enable automated infographic design,
which has been widely used in commercial software,
such as Microsoft PowerPoint and Adobe Illustrator. 
These systems can automatically generate infographics
by plugging in data to a design template.
Although easy to use,
these systems typically only provide limited types of templates with default styles,
which leads to a lack of diversity.
By contrast,
many infographics ``in the wild'' with diverse styles can only be accessed as images in a bitmap format.
If users want to follow the styles of these bitmap infographics,
they have to manually create their own infographics,
which is difficult and tedious.

In this work,
we investigate the methods of automatically 
extracting an extensible template from a bitmap infographic.
Compared to \emph{editable} templates,
\emph{extensible} templates contain not only the editable elements
but also the semantic roles of these elements,
which enable the automatic extension with updated data.
Previous works~\cite{Poco2017, Poco2018} attempt to extract visual encodings
and color mappings from chart images based on rules and machine learning (ML) methods, 
by utilizing the legends, axes, plot areas, and common layouts.
However, the content of infographics can be unstructured and manifold.
This makes it challenging to analyze infographic images and extract extensible templates from them.
As a first step towards automated infographic design,
we focus on the timeline infographics,
which have been widely used for centuries 
and whose design space has been extensively studied~\cite{Brehmer2017}.

Automatically extracting an extensible template from a bitmap timeline infographic is non-trivial.
Particularly, two obstacles stand in the way.
First, it is challenging to interpret a bitmap timeline infographic automatically.
Understanding the content of the infographic is 
necessary for automating the extraction.
However, the elements in an infographic can be distributed in any place 
with any \cmo{style} (\emph{e.g.}, shapes, colors, and sizes, \emph{etc.})
It is difficult for a machine to interpret the infographic that can only be accessed in pixels.
Second, 
it is intricate to convert a bitmap infographic to be extensible automatically.
An understanding of a timeline infographic is not enough for using it as a template.
Even if the machine has already obtained structural information of the timeline (\emph{e.g.}, type, orientation, and categories and locations of its elements),
how to convert the timeline to an extensible template remains unclear,
not to mention the information could be incorrect.

To address these challenges, 
we propose a novel end-to-end approach
for automatically extracting an extensible template 
from a bitmap timeline infographic.
Our approach adopts a deconstruction and reconstruction paradigm.
We address the first challenge at the deconstruction stage.
We propose a multi-task deep neural network (DNN) that simultaneously parses 
two kinds of information from a timeline image: global and local information.
Global information includes the \emph{representation}, \emph{scale}, \emph{layout}, and \emph{orientation} of the timeline.
Local information includes the \emph{location}, \emph{category}, and \emph{pixels} of each visual element on the timeline.
These two kinds of information provide a panorama of the timeline.
We tackle the second challenge at the reconstruction stage.
By utilizing the deconstruction results,
we propose a pipeline with three techniques, \emph{i.e.}, \emph{Non-Maximum Merging}, \emph{Redundancy Recover}, and \emph{DL GrabCut},
to extract an extensible template from the infographic.
The output can be used to generate new timelines with updated data.

To evaluate our approach,
we synthesize a timeline dataset with 4296 labeled images
and collect a real-world timeline dataset from \cmo{the Internet}.
We report quantitative evaluations of the two stages over the two datasets. 
We then present examples of automatically extracted templates with various visual styles
and timelines automatically generated 
based on these templates
to qualitatively demonstrate the eperformance.
The results confirm that our approach can effectively 
extract extensible templates from real-world timeline infographics.
Finally, we discuss lessons learned and future opportunities.

Our primary contribution is an automated approach
to extracting extensible templates from bitmap infographic timelines.
The approach consists of 1) a multi-task DNN that automatically deconstructs bitmap timeline infographics 
and 2) a pipeline that automatically reconstructs extensible templates.
We evaluate our approach with quantitative evaluations 
and qualitatively demonstrate its effectiveness with examples.

\section{Related Works}
This section introduces prior studies that are most relevant to our work, including automated visualization design, computational interpretation of visualization, 
and deep learning-based object detection.

\subsection{Automated Visualization Design}
Automated visualization design systems 
aim at producing visual encodings for given input data based on both the criteria summarized by experts (\emph{e.g.}, Bertin~\cite{Bertin1983}, Cleveland and McGill~\cite{McGill1984}) and constraints defined by users~\cite{Moritz2019}.
Prior work on automated visualization design can be classified into two general categories 
based on how the criteria are derived: rule-based and learning-based approaches.

Mackinlay's APT system~\cite{Mackinlay1987} is a pioneering example 
that enumerates, filters, and ranks visualizations using expressiveness and perceptual effectiveness criteria. 
It was extended by SAGE~\cite{Roth1994}, BOZ~\cite{CASNER1991}, and ShowMe~\cite{Mackinlay2007} with additional considerations 
of data properties, low-level perceptual tasks, and candidate groupings.
Recent systems like Voyager and Voyager 2~\cite{Wongsuphasawat2016, Wongsuphasawat2017} 
have further recommended data transformation (\emph{e.g.}, normalization) in addition to visual encodings.
 
Foregoing explicit rules, 
researchers have recently designed learning-based systems that directly learn visualization designs from visualization corpora.
DeepEye~\cite{Luo2018} applies ML models and design rules 
to determine whether a visualization is ``good'' or ``bad''
and recommends the ``good'' candidates.
Data2Vis~\cite{Dibia2018} uses a Recurrent Neural Network 
to automatically translate 
JSON-encoded datasets to Vega-lite~\cite{Satyanarayan2018} specifications.
Draco~\cite{Moritz2019} learns weights between hard and soft constraints that represent users' requirements and design guidelines.
VizML~\cite{Hu2018} trains a fully-connected neural network 
to predict design choices based on input data.
Although we also aim for automated design,
these systems, however, cannot be adapted to infographics.
They focus mainly on recommending \emph{visual encodings} for the input data
(\emph{e.g.}, how to encode data using visual channels).
By contrast,
designing infographics requires additional attention to \emph{visual styles} 
(\emph{e.g.}, how to embellish the visualization with shapes and icons), 
which are omitted in these systems.
In this regard, 
our work is inherently different from them.

\subsection{Computational Interpretation of Visualization}
Computational interpretation of visualization seeks
to enable machines to understand the content of visualization images
(\emph{e.g.}, data, styles, and visual encodings).
According to the targets, 
prior methods can be divided into two categories:
for charts and for infographics.

A general pipeline when interpreting a chart is first to
identify the type of the chart via classifications,
then detect elements (\emph{e.g.}, marks or text) in the chart,
and finally extract the underlying information (\emph{e.g.}, data or visual encodings).
As a pioneer, Savva et al. introduced ReVision~\cite{Savva2011},
in which the graphical and textual features 
are fed into a support vector machine (SVM) model for a chart type classification.
ReVision then localizes the marks and extracts data from pie and bar charts
by using a carefully designed multi-steps method based on image processing and heuristics. 
Siegel et al.~\cite{Siegel2016} extended the method of ReVision to handle line charts. 
They developed a convolutional neural network (CNN) for the chart classification and designed a heuristic approach 
to use legend information for data extraction.
Recently Kafle et al.~\cite{Kafle2018} have used a deep dual-network model 
to directly parse the data from bar charts without heuristic rules.
Instead of extracting the data of charts,
Poco and Heer~\cite{Poco2017}
aimed
to recover the visual encodings.
To complete the task successfully,
they proposed a state-of-the-art approach to interpreting the text in a multi-stage pipeline,
which combines ML and heuristics methods.
Building on this, Poco et al.~\cite{Poco2018} further explored the color mapping extraction of visualization images.

Apart from charts,
researchers have explored the computational interpretation of infographics.
Bylinskii et al.~\cite{Bylinskii2017a} used fully convolutional networks (FCNs) to predict 
the visual saliency of an infographic.
Bylinskii et al.~\cite{Bylinskii2017} also applied DNNs
to select representative textual and visual elements from an infographic automatically. 
On the basis of several deep learning models, 
Kembhavi et al.~\cite{Kembhavi2016} designed a multi-stage 
approach to parse the relationships among elements 
in diagrams in science textbooks.
\cmo{
    More recent research investigated using DNNs to
    detect UI components in mobile apps~\cite{Liu2018}
    and icons in infographics~\cite{Madan2018}.
}
Although these methods enable
computational understanding of an infographic from certain perspectives,
the information they interpret cannot be used to reconstruct an extensible template 
(\emph{e.g.}, how to change or extend the content of an infographic is unknown).
We take a first step towards the interpretation of infographics 
for an automated design purpose.
Unlike using multiple models and handcrafted features,
our approach uses one end-to-end DNN to complete the interpretation.

\subsection{Deep Learning-based Object Detection}
To extracting an extensible template,
we need to understand each object on it.
We achieve this goal with deep learning-based object detection.
Object detection is a computer vision \cmo{(CV)} task 
whose goal is to localize each object using a bounding box (\emph{i.e.}, \emph{where}) and classify its category (\emph{i.e.}, \emph{what}).
Deep learning-based object detection methods can either be one-stage~\cite{Redmon2015, Liu2016, Lin2017b}
or multi-stage~\cite{Girshick2012, Girshick2015, Ren2015, KaimingHe2017}.
One-stage models directly predict objects' bounding box and category
without involving intermediate tasks.
YOLO~\cite{Redmon2015} is a representative one-stage model
that divides the image into small cells 
and predicts bounding boxes for each cell.
One-stage models have the advantage of fast detection in real time,
which affects accuracy.
By contrast,
multi-stage models can predict accurately, but are often less time efficient.
Multi-stage models, such as RCNN~\cite{Girshick2012},
usually first propose a manageable number of candidate regions (\emph{region proposals}) that may contain objects.
If an object exists within, then they will predict its bounding box and category.
Time consumption is not our first priority,
so we base our work on a multi-stage model.
Mask R-CNN~\cite{KaimingHe2017} is a leading multi-stage model in several benchmarks.
It can further predict the pixels of an object within its bounding box (\emph{i.e.}, \emph{Instance Segmentation}).
We extend Mask R-CNN to interpret 
not only the information of objects (\emph{i.e.}, local)
but also that of the entire timeline infographic (\emph{i.e.}, global).
To the best of our knowledge,
we are the first to adopt this kind of instance segmentation networks
to deal with the infographics interpretation problem.

\section{Problem Statement}
This section introduces
the background of timeline infographics and the problem,
overview of the proposed approach,
and the datasets.

\subsection{Background}

\begin{figure}[h]
\vspace{-2mm}
    \centering 
    \includegraphics[width=0.9\columnwidth]{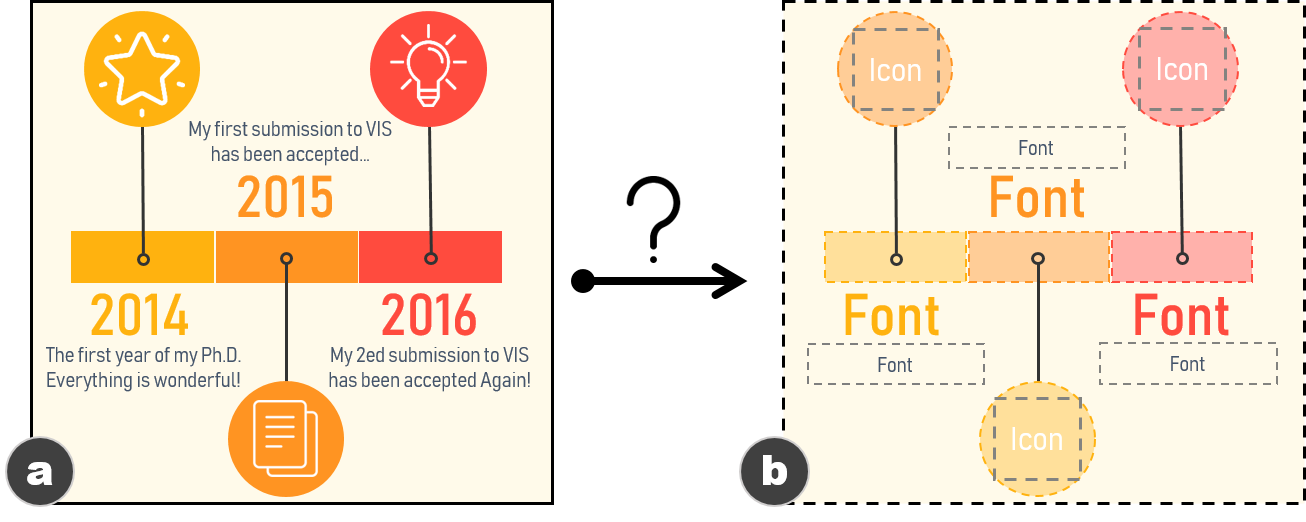}
    \caption{Given a bitmap timeline infographic, we seek to extract its extensible template automatically.}
    \label{fig:problem}
\vspace{-2mm}
\end{figure}

Timeline infographics have been recently investigated by Brehmer et al.~\cite{Brehmer2017}.
We briefly describe the \cmo{insights from them} as follows:
\begin{itemize} [leftmargin=*]
	\item \underline{\emph{Timeline Data.}} 
	A timeline presents interval event data (\emph{i.e.}, a sequence of events),
	which is different from 
	continuous quantitative time-series data. 
	A timeline infographic for storytelling usually has a small underlying dataset
	because the storyteller is assumed to have already distilled 
	the narrative points from the raw dataset.

	\item \underline{\emph{Timeline Design.}}
	A timeline can be described \cmo{as a combination of three dimensions,
	namely, \emph{representation}, \emph{scale}, and \emph{layout}.
    The combination can be used as the \emph{type} of timelines.}
	No more than five options are available for each dimension (\autoref{table:tl_dimensions}).
    \cmo{Besides, only 20 out of 100 combinations of these options are viable.}
    These dimensions indicate how the events are organized in a timeline.
	For example,
	events are placed along a straight line
	in a \emph{linear} representation,
	which is the most common way to represent a timeline.
	Typically, an event is visually encoded by a graphical mark, such as the rectangles in \autoref{fig:problem}.
	The position of this mark is used to encode the occurred time of the event.
	Extra annotations (\emph{e.g.}, text or icons) are added, 
	commonly adjacent to the event mark,
	to depicts the details of an event.
\end{itemize}

\begin{table}[h]
    \vspace{-2mm}
	\caption{The design dimensions to depict a timeline from~\cite{Brehmer2017}}
	\centering
	\label{table:tl_dimensions}
	\begin{tabular}{l@{\hskip 4pt}l}
	  \toprule
	    \multicolumn{2}{c}{\textbf{Design Options}} \\ \midrule
		\textbf{Representation} & Linear, Radial, Grid, Spiral, Arbitrary \\
		\textbf{Scale}  & Chronological, Relative, Logarithmic, \\ 
		      			& Sequential, Sequential + Interim Duration  \\ 
		\textbf{Layout} &  Unified, Faceted, Segmented,  \\ 
						& Faceted + Segmented \\
	  \bottomrule
    \end{tabular}
    \vspace{-2mm}
\end{table}

In practice, infographic timelines are widely spread in the form of bitmap images.
However, they are not easy to reproduce.
Given a bitmap timeline,
we aim to extract its extensible template (\autoref{fig:problem}) automatically.
To this end,
two requirements should be fulfilled:
\begin{itemize} [leftmargin=*]
	\item \textbf{Parse the content.}
	The machine should first parse the content of the image.
    A computational understanding of an image 
    can be represented as a structural information, 
    which is necessary for an automation process.
    However, the infographic image can only be accessed in pixels, 
    which is a byte array with the shape of $width \times height \times RGB$.
    A process is required to take the bitmap image as input
    and output its structural information.

	\item \textbf{Construct the template.}
	With the structural information of the image as a basis,
	the machine should be able to
	construct 
	an extensible template out of it automatically.
	The template should contain detail information (\emph{e.g.}, position, color, font, and shape) of the elements to be reused 
	and the elements to be updated.
	Given the image and its structural information,
	another process should be involved to extract such types of detail information.
\end{itemize}

\subsection{Approach Overview}
To fulfill the two requirements above,
we design a two-step approach, 
starting from defining the input and output of each step.

\textbf{Deconstruction.}
The goal of the first step (\autoref{fig:teaser}a and \autoref{fig:teaser}b) is to parse structural information from the input, a bitmap timeline infographic $I$.
For the output,
we define two kinds of information,
namely, the global one $G$ and the local one $L$.
The global information is about the entire timeline,
including its 
\cmo{three dimensions mentioned in~\autoref{table:tl_dimensions}}
\cmo{and its} \emph{orientation}.
The local information is about each individual element,
including its category (\emph{what}), location (\emph{where}), and the pixel-wise mask (\emph{which pixels}).
Therefore, the ideal process of the first step can be formulated as
a mapping function $f$:
\vspace{-1mm}
\begin{equation}
    f: I \rightarrow (G, L)
\end{equation}
\vspace{-4mm}

We propose to
approximate $f$ using a DNN model $h \approx f$
with a set of parameters $\Theta$.
This set of parameters $\Theta$ can be learned from a corpus 
$\mathbb{C} = \{(I_i: (G_i, L_i))\}_{i=1}^n$,
where each entry $(I_i: (G_i, L_i))$ is a bitmap image 
associated with its global and local information.
Hence, we can obtain the output via $(G, L) = h(I | \Theta)$.

\textbf{Reconstruction.}
To reconstruct the extensible template,
a function $g$ should take the bitmap infographics $I$
and its global and local information $G$, $L$ as the input,
and return the detail information about
elements to be reused $E_r$ (\emph{e.g.}, the rectangle and circle marks in \autoref{fig:problem}a)
and elements to be updated $E_u$ (\emph{e.g.}, the text and icons in \autoref{fig:problem}a), \emph{i.e.},
\vspace{-1mm}
\begin{equation}
    g: (I, G, L) \rightarrow (E_r, E_u)
\end{equation}
\vspace{-4mm}

$E$ is a set of elements,
each of which is represented as a set of attributes, 
\emph{i.e.}, $E = \{ \textbf{$e$}^i := (a^1, a^2, ..., a^m)\}_{i=1}^n$.
According to $G$ and $L$, 
we can infer attributes of elements in $E_r$ and $E_u$,
such as \emph{size}, \emph{shape}, \emph{color}, 
\emph{position}, and \emph{offset} to others.
We highlight the necessary attributes for enabling extensible templates.
For $E_r$, 
the essential attribute is the graphical marks to be reused (\emph{e.g.}, the rectangle marks in \autoref{fig:problem}a).
Hence, we need to segment the pixels of $E_r$ from the original image.
As for $E_u$,
the attributes related to the \emph{font} (\emph{e.g.}, \emph{font family}, \emph{size}, \emph{color}, \emph{etc.})
must be identified
to maintain the styles of the updated content.
In addition, we note that the outputs from $h$ may not be perfect,
reducing the quality of the outputs of $g$.
Thus, $g$ should be smart enough to correct errors in $G$ and $L$
as much as possible.

Considering these issues,
we design a heuristic-based pipeline, with three novel techniques,
as $g$ to automatically output $E_r$ and $E_u$.

\subsection{Datasets}

\begin{figure}[h]
\centering 
\includegraphics[width=0.98\columnwidth]{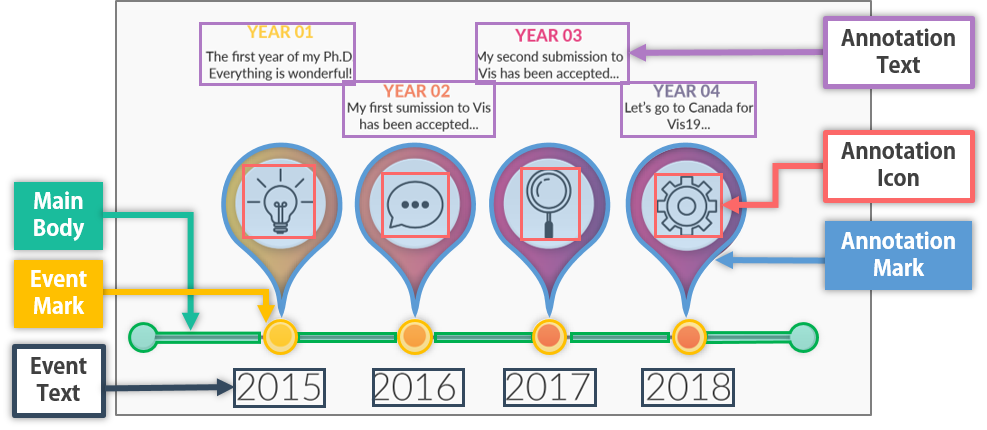}
\caption{Categories of elements in a timeline infographic. The \emph{event mark}, \emph{annotation mark}, and \emph{main body} can be reused, while others need to be updated.}
\label{fig:labels}
\vspace{-2mm}
\end{figure}

\begin{figure*}[thb]
    \centering 
    \includegraphics[width=2\columnwidth]{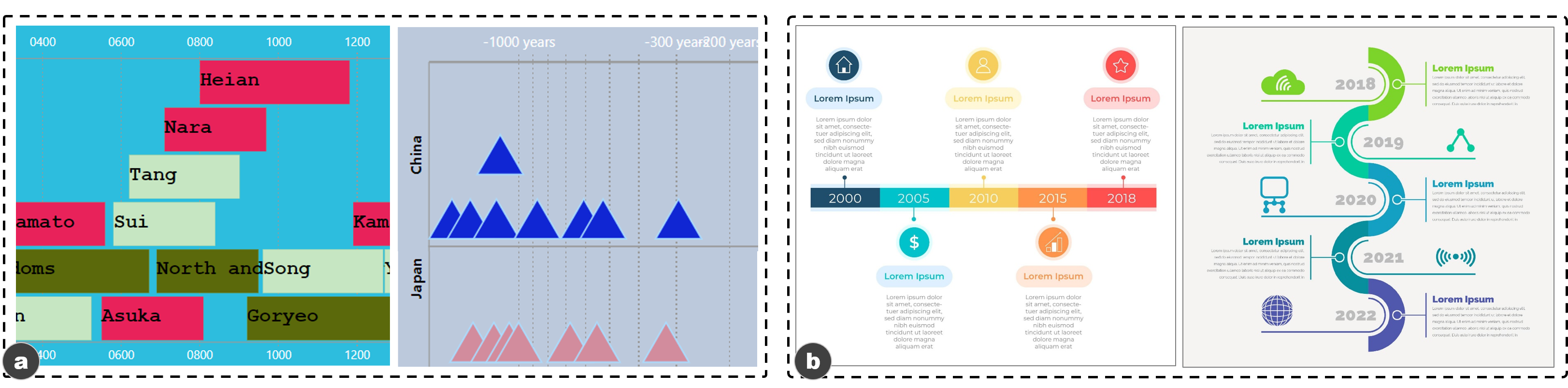}
    \caption{Example timelines from:
    a) a synthetic dataset $D_1$, which shows two different scales,
    and b) a real-world dataset $D_2$, which shows two different orientations.}
    \label{fig:dataset}
    \vspace{-4mm}
\end{figure*}

We use two datasets to train the model $h$ and evaluate our approach.
The first one (referred to as $D_1$) is a synthetic dataset.
We extended TimelineStoryteller (TS)~\cite{timelineStoryteller}, 
a timeline authoring tool,
to generate $D_1$, covering all types of timelines.
The second dataset (referred to as $D_2$) consists of real-world timelines, collected from Google Image~\cite{googleImage}, Pinterest~\cite{pInterest}, and FreePicker~\cite{freePik} by using the search keywords \emph{timeline infographic} and \emph{infographic timeline}.
$D_2$ has more diverse styles, especially for marks,
and it covers the most common types of timelines.
The resolutions of images are in the range of $[512, 3880] \times [512, 4330]$.
To scope this work,
we focus on timelines that
have less than 20 events
and whose events have the same number and types of annotations (\emph{e.g.}, text and icon). 
We also exclude the titles, footnotes, and legends. 

\begin{table}[h]
	\vspace{-2mm}
	\caption{The number of annotations per category of each dataset.}
	\centering
	\label{table:dataset}
	\addtolength{\tabcolsep}{-2pt}
	\begin{tabular}{l@{\hskip 0in}cccccc}
	  \toprule
	  \multirow{2}{1.2cm}{Dataset} & \#Event & \#Event & \#Annot. & \#Annot.  & \#Annot.  & \#Main  \\
	    & Mark & Text & Mark & Text & Icon & Body  \\ \midrule
		$D_1$ & 83498 & 61324 & 4030 & 60036 & - & - \\
		$D_2$ & 2318 & 2305 & 2227 & 2937 & 1497 & 1340 \\ 
	  \bottomrule
    \end{tabular}
    \addtolength{\tabcolsep}{2pt}
    \vspace{-2mm}
\end{table}

\textbf{Collection.}
For $D_1$,
TS
allows us to generate timeline images with various visual encodings and styles.
We generated timeline images using nine embedded datasets of TS 
to cover the design space of timelines.
To increase diversity,
we randomly modified 
the timeline orientation,
the style of graphical marks (including color, size, and shape), 
texts (font, size, color, offset to others),
and the background (color)
in a curated range that guarantee the viability of the timeline.
We created 9592 timelines in this process.

For $D_2$,
we implemented crawlers to download the search results.
The crawling process was manually monitored
and stopped when 10 consecutive return results are not timelines.
We collected 1138 timelines in this process.
Following, four of the coauthors separately reviewed all the timelines
to remove the repeated and problematic instances, 
such as images with heavy watermarks or with low resolutions (\emph{i.e.}, smaller than $512 \times 512$),
\cmo{and timelines that out of the scope of this work}.
They \cmo{obtained} 412 \cmo{remaining} timelines.
The scale of $D_2$ is consistent with 
 manually collected visualization datasets in similar research~\cite{Poco2017, Poco2018, Brehmer2017}.
Among the five representations 
in~\autoref{table:tl_dimensions},
\emph{radial}, \emph{grid}, or \emph{spiral} representations appear 
only 19/412 (4.6\%) timelines, 
whereas the rest 393 timelines are with \emph{linear} or \emph{arbitrary} representations.
This ratio is consistent with \cite{Brehmer2017} (23/263, 8.7\%).
Considering the scarce number of the \emph{radial}, \emph{grid}, or \emph{spiral} representations,
we excluded them in $D_1$ and $D_2$
and focused on the more common \emph{linear} and \emph{arbitrary} representations.

\textbf{Labeling.}
To identify the categories of elements in a timeline,
four of the coauthors independently reviewed all the timelines in $D_1$ and $D_2$.
Each of them iteratively summarized a set of 
mutually exclusive categories that can be used to depict elements in a timeline infographic.
Gathering the reviews resulted in six categories (\autoref{fig:labels}).
We explain the details of these categories in the supplemental material.

Each timeline in $D_1$ was then
converted from SVG to bitmap format
and annotated with its representation, scale, layout, and orientation.
We also analyzed the SVG and the bitmap 
to generate the annotations for each element in a timeline,
including its category (from the label sets in \autoref{fig:labels}),
bounding box (referred to as \emph{bbox}),
and pixel-wise mask (referred to as \emph{mask}).
For each timeline in $D_2$,
we manually annotated its representation, scale, layout, and orientation,
as well as the category, bbox, and mask of each element,
by using our annotation tool that is built on Microsoft PowerPoint.
Finally, $D_1$ contains 4296 timelines,
whereas $D_2$ contains 393.
Figure \ref{fig:dataset} and \autoref{table:dataset} 
present samples and statistics of these timelines, respectively.

\section{Deconstruction}

Parsing bitmap timeline infographics 
to extract structural information 
is difficult 
due to the absence of fixed rules for 
the styles and layouts of timeline elements.
We achieve this goal
from two perspectives, global and local.
In contrast with prior studies~\cite{Poco2017, Siegel2016}
that extracted structural information from charts
using different methods in multiple steps,
we use
a DNN to extract structural information in one shot.


\subsection{Parsing Global Information}
\label{sec:understand_global}

Our dataset comprises 10 types of timelines.
The \emph{type},
\cmo{\emph{i.e.}, the combination of the three dimensions in~\autoref{table:tl_dimensions},}
\cmo{is} necessary for constructing an extensible template.
In addition to the \emph{type} of timeline,
the \emph{orientation},
which could be \emph{horizontal}, \emph{vertical}, and \emph{others},
is equally indispensable for the template.
As \emph{type} and \emph{orientation} only involve a few discrete choices, 
we can identify them through classification.

Taking into account that CNN models have shown excellent capability in chart classification~\cite{Poco2017, Siegel2016, Jung2017}, we propose a CNN-based classifier to recognize the \emph{type} and \emph{orientation} of a timeline.

\begin{figure}[h]
	\centering 
	\includegraphics[width=\columnwidth]{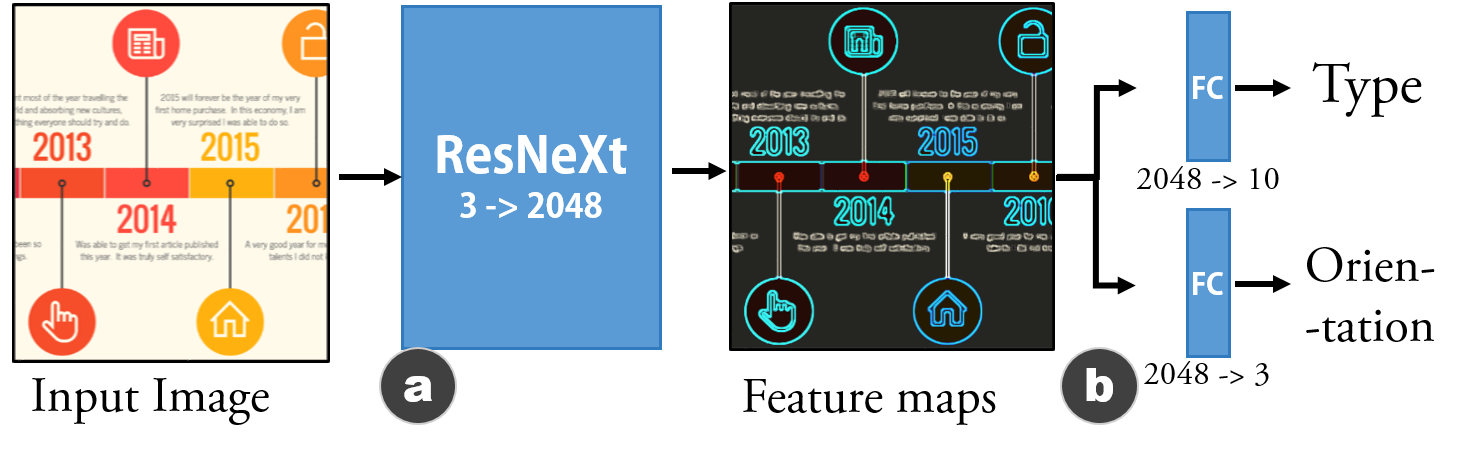}
	\caption{Initial architecture to parse the global information. 
	After extracting the feature map of an image, 
	two FC layers are used to classify its \emph{type} and \emph{orientation}.}
	\label{fig:init_archit}
	\vspace{-2mm}
\end{figure}

Many CNN architectures have been proposed (\emph{e.g.}, AlexNet~\cite{Krizhevsky2012}, GoogLeNet~\cite{Szegedy2015}).
\cmo{We use ResNeXt~\cite{Xie2017} (\autoref{fig:init_archit}a)
to extract the features of a timeline infographic,
since ResNeXt achieves state-of-the-art performance in many \cmo{CV} tasks.}
\cmo{It takes a 3-channel image (\emph{i.e.}, RGB) as input 
and output a feature map with 2048 channels.}
We then use two siblings fully connected (FC) layers (\autoref{fig:init_archit}b)
as Class heads
to predict the timeline's \emph{type} 
and \emph{orientation} based on the feature map. 

\subsection{Parsing Local Information}
After parsing the global information,
the machine should further extract the local information of the timeline.
We have defined six categories of elements (\autoref{fig:labels}) in a timeline.
We need to detect each element in the timeline (where and what) 
and segment it from others (which pixels).

To tackle these tasks,
a possible solution is to 
solve them one by one using well-established methods.
For example,
we can use sliding windows~\cite{Lampert2018}
to localize elements,
\cmo{then employ SVM}
to determine the category of the element within,
and lastly segment the element from the image.
This multi-step solution can be effective 
and has been used in previous works~\cite{Poco2017, Siegel2016, Jung2017, Kembhavi2016}.
However, given the ad-hoc nature, 
extending this solution to other scenarios is challenging.
Therefore, we prefer to adopt a unified method to complete all tasks.

\begin{figure}[h]
\centering 
\vspace{-2mm}
\includegraphics[width=\columnwidth]{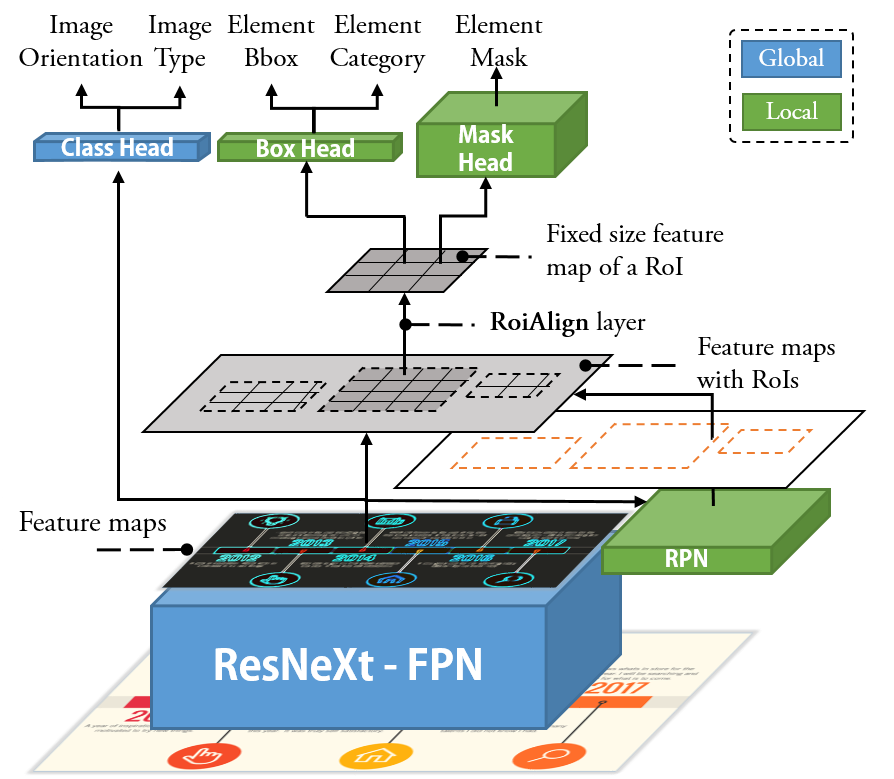}
\caption{Complete architecture to parse both global and local information simultaneously.
Apart from the components in \autoref{fig:init_archit},
we add more components (in green color) to parse local information.
}
\label{fig:architecture}
\vspace{-2mm}
\end{figure}

Considering that we have already extracted the feature maps of the infographic
in \autoref{sec:understand_global},
we propose to reuse these feature maps,
which contain rich information of the image.
Specifically, we extend the classification model in~\autoref{fig:init_archit}
by adding components for object detection to parse the local information.
\cmo{We achieve this extension using} Mask R-CNN~\cite{KaimingHe2017}, 
\cmo{a leading architecture that can detect objects and predict their pixel masks.}
\cmo{By this means,}
our model can simultaneously finish all five tasks
\cmo{(\emph{i.e.}, two global and three local)}
in one shot.
The complete architecture is depicted in~\autoref{fig:architecture}.
We successfully train this multi-task learning model 
and achieve a good performance.

In the architecture (\autoref{fig:architecture}),
we first extend ResNeXt with Feature Pyramid Network~\cite{Lin2017a} (FPN).
FPN is a top-down architecture 
\cmo{that} can build semantically strong feature maps 
at multiple scales 
using the feature maps from ResNeXt.
FPN makes our model scale-invariant 
and able to handle images of vastly different resolution.
We then feed the feature maps from the ResNeXt-FPN
into a Region Proposal Network~\cite{Ren2015} (RPN) to localize elements in a timeline.
RPN is an FCN that simultaneously predicts element locations (\emph{i.e.}, by bbox)
and objectness scores (\emph{i.e.}, whether there is an object within the bbox) in an image.
These element location hypotheses
are then be used to extract 
regions of interest (RoIs) 
from the feature maps.
Each RoI is normalized to a fixed size using a RoIAlign layer
and then passed to two heads, namely, a Box head and a Mask head.
The Box head uses two sibling FC layers 
to classify the category and regress the bbox of the element.
The Mask head uses an FCN for predicting the pixels of the element within the bbox.
Additional details on the architecture and training process are presented 
in the \cmo{supplemental material.}

\subsection{Validation}
\label{sec:deconstruct_validation}
Our model is implemented using Pytorch~\cite{pytorch}
with 
two types of CNN backbone,
namely, ResNeXt-50 (R50) and ResNeXt-101 (R101), 
following the standard configurations~\cite{Xie2017}.
R50 has 50 layers,
which is more lightweight and easier to train,
while R101 has 101 layers,
which performs better in \cmo{CV} tasks at the cost
of efficiency and is more difficult to train.
We trained these two implementations of our model using $D_1$ and $D_2$ together.
We randomly split the images in $D_1$ and $D_2$ into $9:1$
such that no testing sample is in the training set. 
To increase the diversity of the training data,
we conduct several data augmentation strategies, 
including random horizontal or vertical flip, 
random \cmo{90-degree} rotations \cmo{(the labels are updated accordingly)},
and random color channels swap.
Finally, the number of training samples for one epoch is 33760.
We evaluated models trained with 10 epochs on the two datasets separately.
We first report the performance of parsing global information
and then report the average precision ($AP$) on parsing local information.
Reported numbers are averaged over 10 independent runs.

\textbf{Parsing Global Information}.
To access the performance of our model on the two classification tasks 
(\emph{i.e.}, 10 classes \cmo{of} timeline \emph{type} and 3 classes \cmo{of} \emph{orientation}),
we calculate the precision, recall and F1-score.
Table \ref{table:global} presents the results.

\begin{table}[h]
	\caption{Classification of timeline types and orientations.}
	\centering
	\label{table:global}
	\begin{tabular}{lcccccc}
	  \toprule
	  \multirow{2}{1.1cm}{Dataset / \ Backbone} & \multicolumn{3}{c}{Type} & \multicolumn{3}{c}{Orientation} \\
	  \cmidrule(l{2pt}r{2pt}){2-4} \cmidrule(l{2pt}r{2pt}){5-7}
		    & {$Pre.\%$} & {$Rec.\%$} & {$F1\%$} & {$Pre.\%$} & {$Rec.\%$} & {$F1\%$}\\ \midrule
			$D_1$ / R50 & 99.1 & 99.1 & 99.1 & 100.0 & 100.0 & 100.0 \\ 
			$D_1$ / R101 & 99.5 & 99.5 & 99.5 & 100.0 & 100.0 & 100.0 \\ 
			$D_2$ / R50 & 88.7 & 86.4 & 87.5 & 97.7 &  97.1 & 97.4 \\ 
			$D_2$ / R101 & 92.2 & 90.9 & 91.5 & 97.7 & 97.1 & 97.4 \\ 
	  \bottomrule
	\end{tabular}
  \end{table}

Both implementations achieve good performance on $D_1$ and $D_2$.
As expected,
R101 has a better performance
on $D_1$ and $D_2$ than R50.
The classification of \emph{type} on $D_2$ performs worse than that on $D_1$,
which is largely due to the more diversity and small size of $D_2$.
Nevertheless, \cmo{F1-score} is still higher than 90\% when using R101.

\begin{wrapfigure}{R}{0.13\textwidth}
	\centering
	\includegraphics[width=0.13\textwidth]{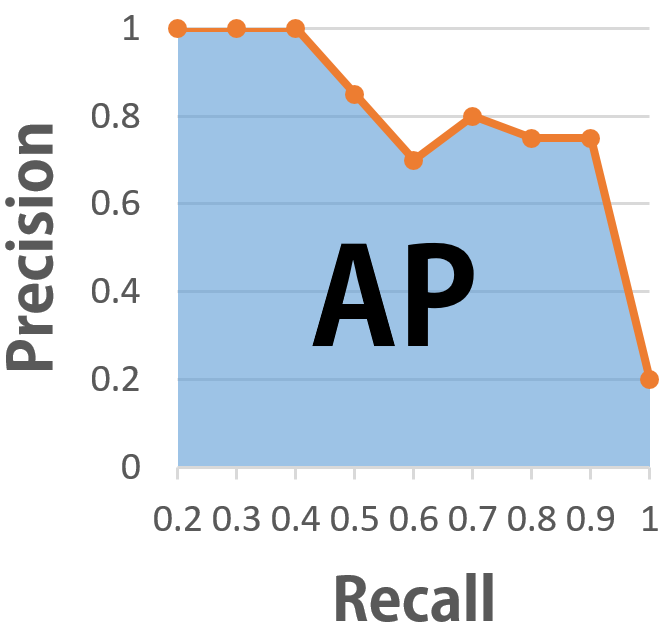}
	\caption{AP.}
	\label{fig:mAP}
\end{wrapfigure}

\vspace{2mm}
\textbf{Parsing Local Information}.
To evaluate the performance of parsing local information,
we use the metrics in COCO challenge~\cite{coco}.
COCO is a large-scale object detection and segmentation dataset
that contains more than 330K images with high-quality annotations.
\cmo{It 
uses \emph{AP} metrics~\cite{Everingham15} to access 
the three tasks (\emph{i.e.}, \emph{what}, \emph{where}, and \emph{which pixels}) together.}
Basically, \emph{AP} is a measure of precision-recall tradeoff
calculated using all possible confidence level
that is represented by the classification score associated with each predicted bbox.
Intuitively, \emph{AP} is the area under the precision-recall curve (\autoref{fig:mAP}).
To calculate the precision and recall at a confidence level,
we first need to calculate the intersection over union (\emph{IoU}) 
between each predicted bbox $B_p$ and its corresponding ground truth $B_gt$
by $IoU = \frac{area(B_p \cap B_{gt})}{area(B_p \cup  B_{gt})}$. 
If the $IoU$ exceeds a threshold (\emph{e.g.}, 0.5),
the prediction is considered as a true positive; otherwise a false positive.
We can then further calculate the precision and recall over all confidence level to draw the curve.

\begin{table}[h]
	\vspace{1mm}
	\caption{Average Precision of parsing local information.}
	\centering
	\label{table:local}
	\begin{tabular}{lcccccc}
	  \toprule
	  \multirow{2}{1.1cm}{Dataset / Backbone} & \multicolumn{3}{c}{BBox} & \multicolumn{3}{c}{Mask} \\
			\cmidrule(l{2pt}r{2pt}){2-4} \cmidrule(l{2pt}r{2pt}){5-7}
		& {$AP_{50:95}$} & {${AP}_{50}$} & {${AP}_{75}$} & {$AP_{50:95}$} & {${AP}_{50}$} & {${AP}_{75}$} \\ \midrule
		$D_1$ / R50 & 79.0 & 93.6 & 88.0 & 79.8 & 96.4 & 91.6 \\ 
		$D_1$ / R101 & 81.9 & 93.9 & 89.1 & 79.9 & 96.9 & 91.1 \\  
		$D_2$ / R50 & 53.4 & 79.3 & 61.8 & 56.9 & 80.1 & 61.6\\ 
		$D_2$ / R101 & 56.4 & 81.7 & 64.9 & 59.1 & 82.5 & 65.1 \\ \midrule 
		$COCO^*$  & 39.8 & 62.3 & 43.4 & 37.1 & 60.0 & 39.4 \\
	  \bottomrule
	\end{tabular}
		  *A state-of-the-art performance on COCO dataset reported by~\cite{KaimingHe2017}.
  \end{table}
 
 \vspace{1mm} 
Table~\ref{table:local} presents the \emph{AP} of our model on the two datasets.
The higher the \emph{AP}, the better it is.
We provide state-of-the-art performance on COCO
reported by He et al.~\cite{KaimingHe2017} as a background,
due to the lack of benchmarks.
${AP_{50:95}}$ is the average \emph{AP} over different \emph{IoU}, from $0.5$ to $95$ with step $0.05$. 
${AP_{75}}$ and ${AP_{50}}$ is the \emph{AP} calculated
at $IoU = 0.75$ and $IoU = 0.5$, respectively. 
The larger the \emph{IoU}, the stricter the metric will be.
As indicated in~\autoref{table:local},
our model achieves high \emph{AP}
on bbox detection
and pixel segmentation on $D_1$.
This result is because 
the overall diversity of $D_1$ is limited,
the size of $D_1$ is big enough in terms of its diversity,
and the auto-generated annotations of $D_1$ are perfect \cmo{for enabling effective learning}. 
By contrast,
$D_2$ has more diversity, \cmo{a} smaller size,
and imperfect annotations in comparison with \cmo{that of} $D_1$,
leading to a \cmo{decrease} in performance.
Nevertheless, our model still achieves an acceptable performance on $D_2$,
considering the state-of-the-art performance on COCO.
We further discuss the annotation perfectness of $D_2$ in \autoref{sec:reconstruct_validation}.

\section{Reconstruction}
\label{sec:reconstruction}

\begin{figure}[ht]
	\vspace{-1mm}
	\centering 
	\includegraphics[width=\columnwidth]{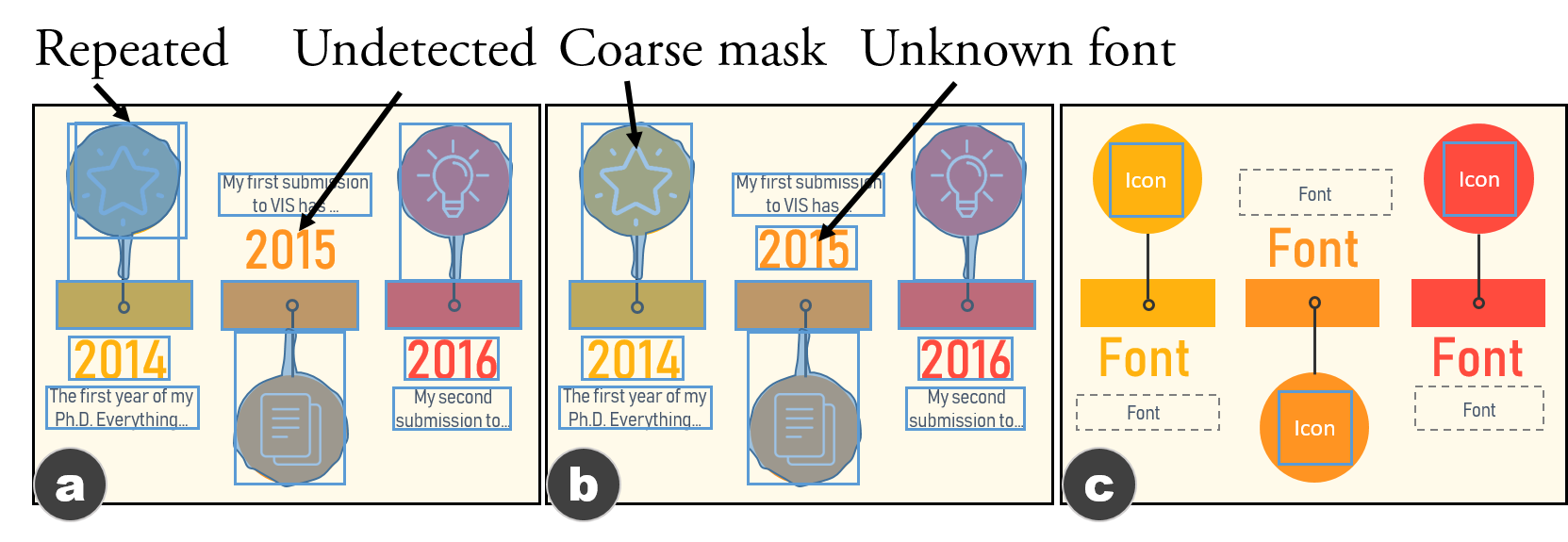}
	\caption{The reconstruction pipeline: 
		a) uses \emph{NMM} and \emph{RR} to eliminate repeated and fix failed detections, respectively;
		b) uses \emph{DL GrabCut} and text recognition to collect the elements to be reused and updated, respectively;
		c) the final outputs can be depicted by a specification.
	}
	\label{fig:rec_pipeline}
	\vspace{-2mm}
\end{figure}

After 
interpreting
a timeline infographic,
the next problem is how to automatically extract an extensible template from it.
We achieve this by
a reconstruction pipeline (\autoref{fig:rec_pipeline}) 
that exploits the outputs from the previous step.
Our pipeline first eliminates the repeated \cmo{bboxes} 
using \emph{Non-Maximum Merging} (\emph{NMM}),
and then infers the missing elements using \emph{Redundancy Recovery} (\emph{RR}).
Next, 
\emph{DL GrabCut} is employed
to 
extract high-quality graphical marks for reuse.
Finally, the font of \emph{event text} and \emph{annotation text} is identified using a publicly available API.
A quantitative validation 
confirms the effectiveness of our pipeline.

\subsection{Eliminate Repeated BBoxes: Non-Maximum Merging}
\begin{figure}[h]
	\vspace{-2mm}
	\centering 
	\includegraphics[width=\columnwidth]{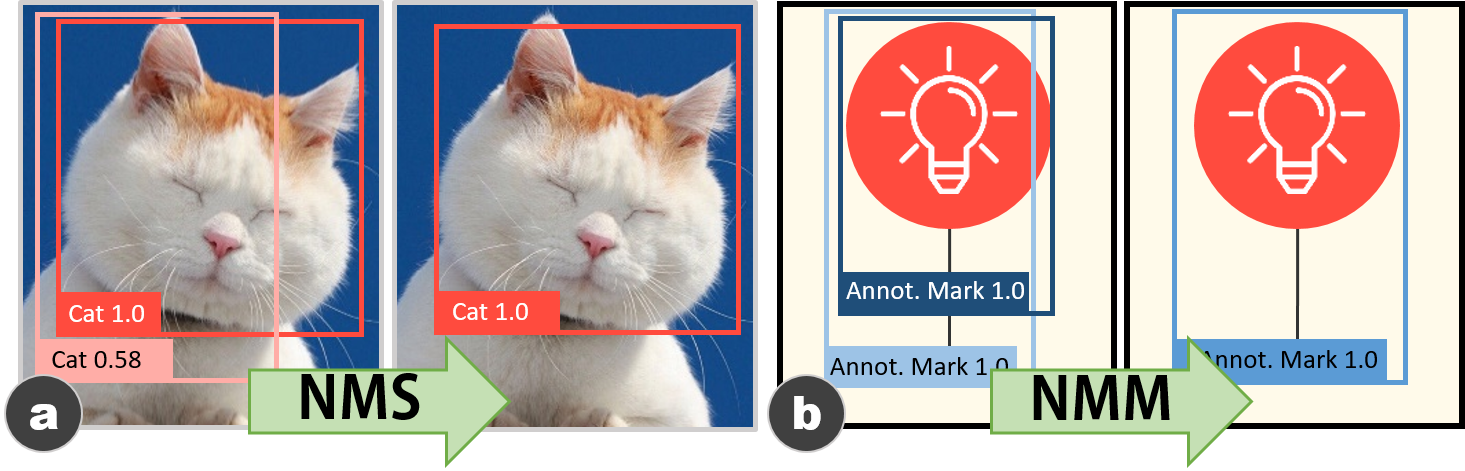}
	\caption{Eliminate repeated bboxes: 
	a) \emph{NMS} keeps the bbox with the highest confidence and removes the others;
	b) \emph{NMM} merges bboxes to the one with the highest confidence and the largest area.}
	\label{fig:nmm}
	\vspace{-2mm}
\end{figure}

Multiple predicted bboxes may exist on one object, 
such as the two bboxes in~\autoref{fig:nmm}a.
\cmo{
A commonly used method to remove repeated bboxes for natural images}
is \emph{Non-Maximum Suppression} (\emph{NMS})~\cite{Girshick2012}.
\emph{NMS} iteratively eliminates bboxes whose confidence score (\emph{i.e.}, the classification score) 
are less than a predefined threshold.
For instance, in~\autoref{fig:nmm}a,
with a threshold of 0.8,
\cmo{\emph{NMS} will eliminate 
the pink bbox with 0.58 
and output the red one.}
However, for infographics,
a part of an object may still be a ``complete'' object,
which hinders the effectiveness of \emph{NMS}.
For example, in~\autoref{fig:nmm}b,
the mark in the steel blue bbox 
and the part of it in the deep blue bbox are both valid annotation marks.
In such case,
each of the two bboxes will be assigned a high confidence score (\emph{e.g.}, 1.0).
Therefore \emph{NMS} cannot eliminate the repeated box.

Therefore,
we design \emph{NMM} to eliminate repeated bboxes.
Specifically,
for \cmo{bboxes} with the same category,
we rank them using the confidence score plus the area (normalize to $[0, 1]$).
For the top 1 bbox,
we merge other bboxes
that overlap with it and exceed a \emph{IoU} threshold
to form a union bbox. 
This process is repeated until all overlapping boxes are merged.
\autoref{fig:nmm}b shows the boxes before and after \emph{NMM}.
In practice, 
we apply both \emph{NMS} and \emph{NMM} separately
and then check the consistency of the shapes 
between the resulted bboxes
and other non-repeated bboxes. 
The most consistent results are kept.

\subsection{Fix Failed Detections: Redundancy Recovery}
\cmo{Our model}
may miss elements (\emph{i.e.}, false negative)
or
detect elements with wrong categories (\emph{i.e.}, false positive). 
To fix these failed detections, 
we leverage the redundant information of timelines
(\emph{e.g.}, each event has the same type of annotations).
Specifically,
for the elements in a timeline,
we first group them along the timeline orientation into clusters,
each of which represents an event.
Then, 
we use the statistics of clusters 
to verify and attempt to fix failed detections.

\textbf{Incorrectly classified elements.}
Some elements in an infographic can be 
classified into wrong categories.
For example, a short annotation text with a fancy font
can be incorrectly classified as an annotation icon.
We adopt a voting mechanism to attempt to fix 
these misclassifications.
For instance,
if more than half of the events contain annotation text,
then an annotation icon,
whose bbox has the same shape as these annotation texts,
of an event should be classified as an annotation text.
Given an event can have multiple annotation texts, 
we restrict that only the annotation texts with the consistent shape of bbox
can vote for each other. This rule is also applied to other categories.

\vspace{1mm}
\textbf{Missing elements.}
We also use a similar voting mechanism to infer the undetected elements.
For example,
in \autoref{fig:rec_pipeline}a,
more than half of the events have an event text.
Thus, for the event without an event text (\emph{i.e.}, the event in 2015),
we assume it should have an event text.
By using heuristic rules,
we can estimate the bbox (\emph{i.e.}, \emph{x}, \emph{y}, \emph{width}, \emph{height}) of its event text
based on those of other event texts.

\subsection{Elements to be Reused}
\begin{figure}[h]
	\centering 
	\vspace{-1mm}
	\includegraphics[width=0.9\columnwidth]{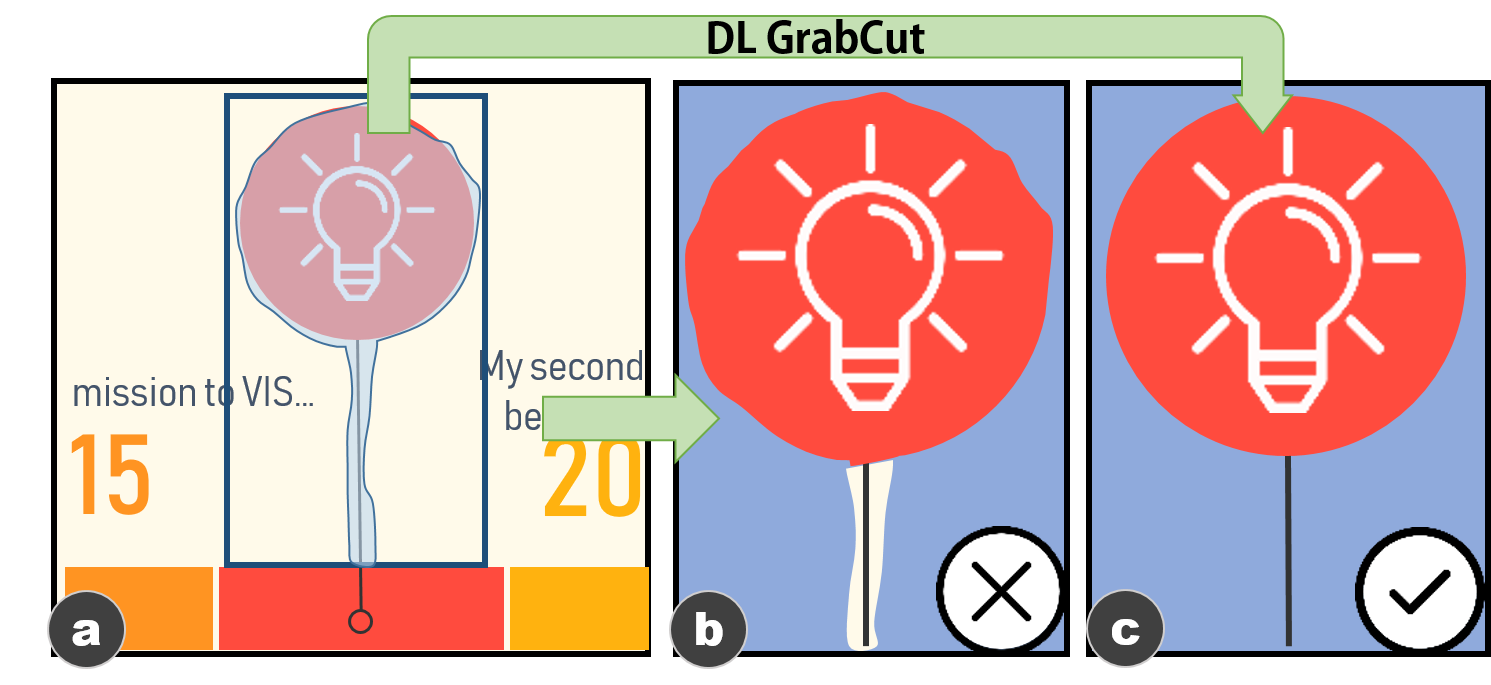}
	\caption{DL model ``interacts'' with GrabCut by using the bbox and mask: 
		a) the bbox and mask predicted by our model;
		b) the predicted mask is coarse;
		c) the refined result from \emph{DL GrabCut}.
	}
	\label{fig:grabcut}
	\vspace{-2mm}
\end{figure}

For an extensible template,
certain elements must be reused
via segmentation from the infographic image.
Our model can predict the pixels (\emph{i.e.}, mask) of each element for the segmentation.
However, the quality of these predicted pixels (\autoref{fig:grabcut}b)
may not be accurate enough for template generation.

To tackle this issue, 
we use the outputs from the model (\autoref{fig:grabcut}a) 
as the input to GrabCut~\cite{Carsten2004} algorithm.
GrabCut is an interactive segmentation algorithm
that has been widely used in production tools, such as  Microsoft PowerPoint.
It 
\cmo{performs well}
especially 
when the background and foreground are not similar
and the edges of the foreground are crisp, 
which is a good fit for our scenario.
To segment an element, 
GrabCut needs a bbox around the target element as an input.
Then, the user can further refine the segmentation results 
by drawing strokes to mark the probable foreground and background area.

Our idea is to automate this process using the outputs of the model
to imitate the user interactions.
For each predicted element,
we use its bbox as the bbox drew by human
and its mask as the user's strokes to refine the segmentation.
By this means,
we leverage the semantic information from the DL model
and the advantage of GrabCut on image processing to 
obtain high-quality masks (\autoref{fig:grabcut}c).

\setcounter{figure}{11}
\begin{figure*}[b]
	\vspace{-2mm}
	\centering 
	\includegraphics[width=2\columnwidth]{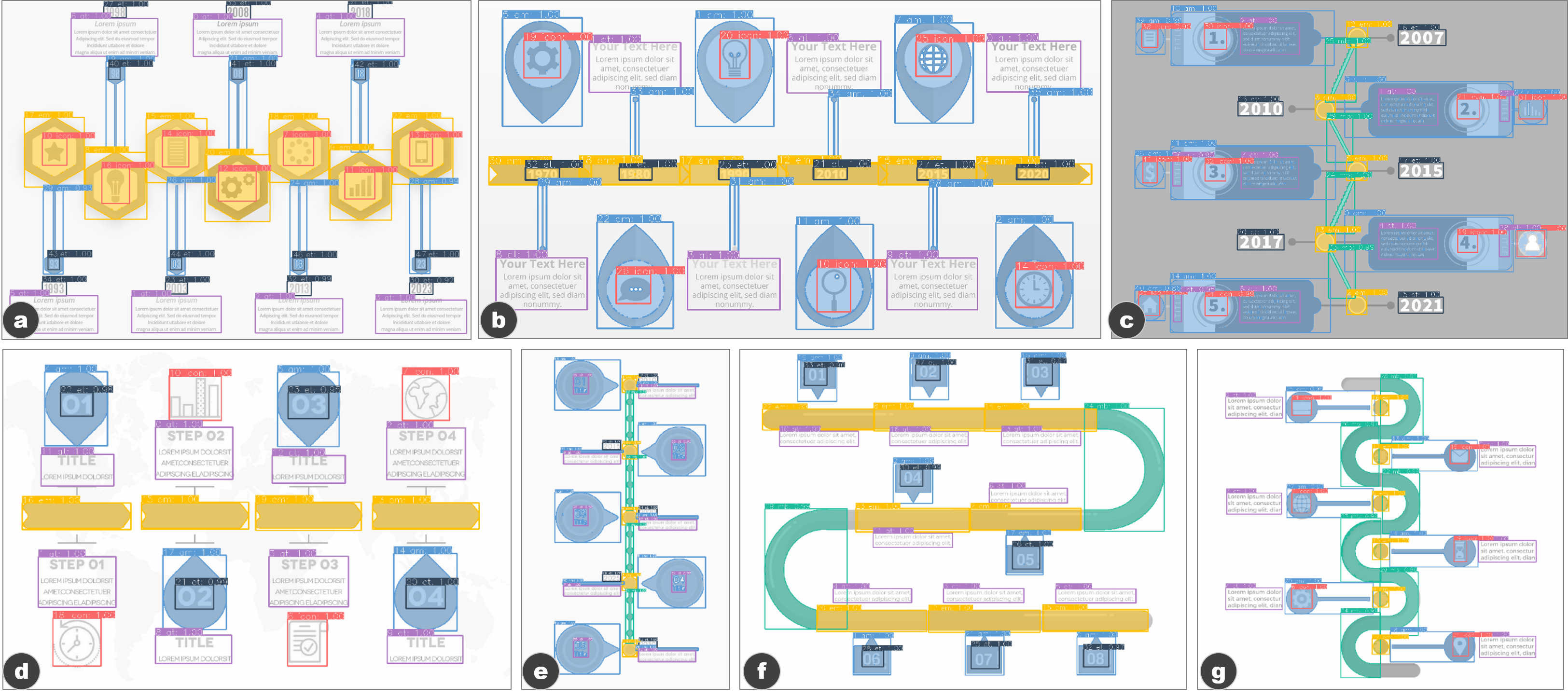}
	\caption{Example results from $D_2$. 
		We visualize the final predicted category, bbox, and mask of each element, 
		following the color legend in~\autoref{fig:labels}.
		We use gray-scale images for a clear demonstration.
		The original images and extra results can be checked in the \cmo{supplemental material}.
	}
	\label{fig:good_cases}
	\vspace{-2mm}
\end{figure*}

\subsection{Elements to be Updated}
Among the six categories of elements,
three categories of elements
need to be updated, namely, \emph{event text}, \emph{annotation text},
and \emph{annotation icon}.
\emph{Annotation icon} can be updated by directly using new icons,
whereas \emph{event text} and \emph{annotation text} should
maintain the same styles with the original infographic,
including their font family, color, and size.
To identify the font family,
we use Font Identifier powered by Fontspring Matcherator~\cite{fontmatcher}.
The font size and color can be calculated and extracted 
from the pixels of the text in the bitmap image.
Some annotation text contains title and body text.
We heuristically identify the text with the larger font size as the title
and the smaller one as the body.
To improve the extensibility of the template,
we further use OCR engine (\emph{i.e.}, Tesseract~\cite{Smith2007}) 
to recognize the content of \emph{event text}
to infer the visual encodings of the timeline \cmo{following the method in~\cite{Poco2017}}.
The final outputs can be depicted using a structural document (\autoref{fig:template}).

\setcounter{figure}{10}
\begin{figure}[h]
	\centering 
	\includegraphics[width=0.98\columnwidth]{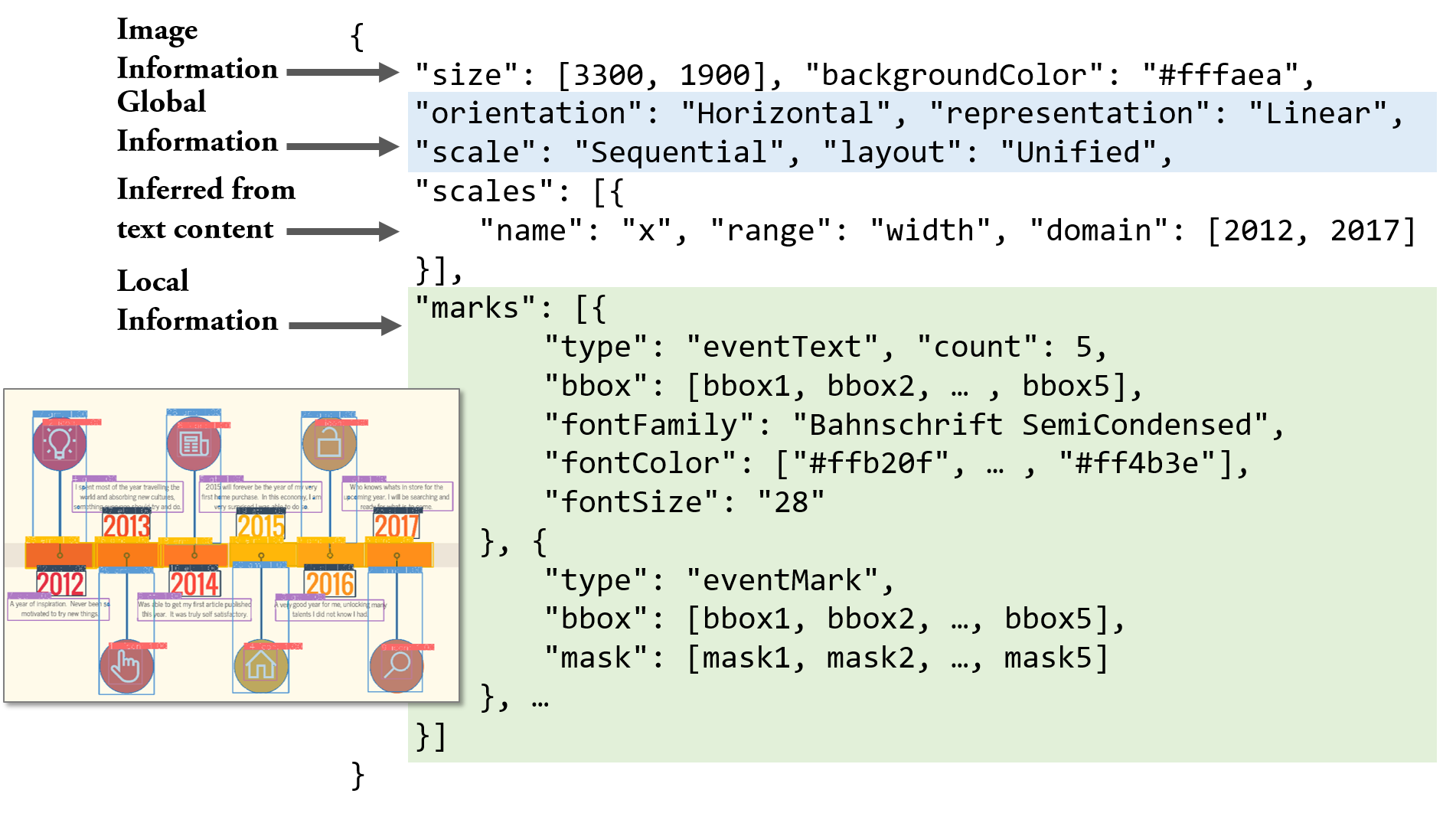}
	\vspace{-2mm}
	\caption{The extensible template can be organized in a \cmo{reusable} document. 
	The \textbf{bbox} is a tuple of $(top, left, width, height)$. 
	The \textbf{mask} is a byte array with shape $width \times height$.}
	\label{fig:template}
	\vspace{-4mm}
\end{figure}

\subsection{Validation}
\label{sec:reconstruct_validation}

To evaluate the effectiveness of our  pipeline,
we reuse the R101 trained in \autoref{sec:deconstruct_validation}.
We are interested in two aspects of our pipeline:
whether it can correct the \cmo{failed detections}
and whether it can \cmo{refine the segmentation results}.
We only test our pipeline on $D_2$,
because the prediction results 
on $D_1$
are good enough to skip the steps 
that we want to access.

To obtain prediction outputs,
we select the confidence level per category 
based on the precision-recall curve calculated in \autoref{sec:deconstruct_validation}.
We calculate the precision and recall of 
the predictions with \emph{IoU} at 0.5 and 0.75.
We then apply our pipeline on the predictions
and calculate the gains of precision and recall on each step.
We expected the following: 
\begin{itemize}
	\item \emph{NMM} can improve the precision of bbox and mask predictions as it removes a few false positives.

	\item \emph{RR} can improve the precision and recall of bbox and mask prediction as it increases the number of true positives.
	
	\item \emph{DL GrabCut} can improve the precision and recall of mask prediction as it improves the quality of masks.
\end{itemize}

Table \ref{table:gain_50} presents 
the gains 
on precision and recall after each step.
The gains at \emph{IoU} 0.5
are close to those at \emph{IoU} 0.75,
which means the gains from the reconstruction pipeline are strict and stable.
As expected, the \emph{NMM} shows a gain on the precision of bboxes and masks predictions.
We also observe a small gain on recall.
The analysis results indicate that such small gain is attributed to
the merging results increasing the number of true positives in some cases 
(\emph{e.g.}, two false positives become one true positive after merging).
Moreover, \emph{RR} shows a gain on the recall of bboxes and masks predictions.
These results confirm that our 
\begin{table}[ht]
	\vspace{1mm}
	\caption{Gains come from Reconstruction at \emph{IoU} 0.5 and 0.75.}
	\centering
	\label{table:gain_50}
	\addtolength{\tabcolsep}{-2.8pt}
	\begin{tabular}{lcccccccc}
	  \toprule
	  \multirow{2}{*}{} & \multicolumn{4}{c}{BBox} & \multicolumn{4}{c}{Mask} \\
			\cmidrule(l{2pt}r{2pt}){2-5} \cmidrule(l{2pt}r{2pt}){6-9}
		& {$Pre_{50}$} & {$Rec_{50}$} & {$Pre_{75}$} & {$Rec_{75}$} & {$Pre_{50}$} & {$Rec_{50}$} & {$Pre_{75}$} & {$Pre_{50}$} \\ \midrule
		Raw & 82.9 & 80.8 & 74.0 & 72.1 & 85.7 & 81.5 & 75.8 & 72.2 \\
		+\emph{NMM} & +2.3 & +1.0 & +2.3 & +0.8 & +1.9 & +0.6 & +1.8 & +0.3 \\
		+\emph{RR} & +1.6 & +2.5 & +1.6 & +2.3 & +2.3 & +2.1 & +2.3 & +2.0 \\
		+\emph{DLGC} & 0.0 & 0.0 & 0.0 & 0.0 & -2.8 & -5.5 & -4.1 & -3.8 \\ \midrule
		Total & 86.8 & 84.1 & 77.9 & 75.4 & 84.0 & 78.4 & 75.9 & 71.0 \\
	  \bottomrule
	\end{tabular}
	\addtolength{\tabcolsep}{2.8pt}
	\vspace{-0.5mm}
  \end{table}

\setcounter{figure}{12}
\begin{figure}[h]
	\vspace{-0mm}
	\centering 
	\includegraphics[width=\columnwidth]{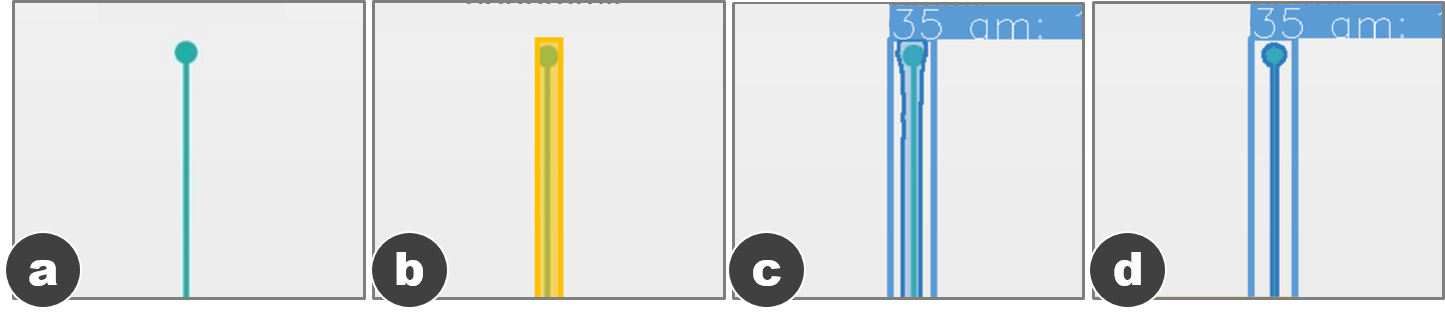}
	\caption{The error from the imperfect label.
		a) an annotation mark and b) its manually labeled mask;
		c) the predicted mask of the annotation mark;
		d) the refined result from \emph{DL GrabCut}.
	}
	\label{fig:mask_example}
	\vspace{-3mm}
\end{figure}

\noindent technique can correct some failed predictions (\emph{i.e} wrong and missing). 

A surprising finding is the decrease in 
the precision and recall of mask predictions.
Our investigation reveals that
this decrease is due to
the imperfect labels.
Figures \ref{fig:mask_example}a and \ref{fig:mask_example}b
show an annotation mark and its label.
Figure \ref{fig:mask_example}c presents the prediction result
and \autoref{fig:mask_example}d shows the result \cmo{refined} by \emph{DL GrabCut}.
The manually labeled mask
encapsulates the graphical mark with empty spaces and a border.
Meanwhile,
the result from \emph{DL GrabCut} perfectly matches
the graphical mark but not the label.
Thus, 
even the result from \emph{DL GrabCut}
is of high quality,
it can also be changed from a true positive to a false positive,
leading to a decrease in precision and recall.
\cmo{
Hence, we manually compare the segmentation results of each element
before and after using \emph{DL GrabCut} 
to verify its effectiveness.
The comparison 
confirms the usefulness of \emph{DL GrabCut} in the pipeline.
}

\section{Extracted Results and Generated Examples}

Example extracted results are shown in \autoref{fig:good_cases}.
High-resolution results can be found in the \cmo{supplemental material.}
Our approach can extract templates 
from not only the timelines with \emph{linear} representations 
(\emph{e.g.}, horizontal \autoref{fig:good_cases}b and \ref{fig:good_cases}d,
and vertical \autoref{fig:good_cases}e and \ref{fig:good_cases}g),
but also those with \emph{arbitrary} representations 
(\autoref{fig:good_cases}a, \ref{fig:good_cases}c, and \ref{fig:good_cases}f).
Figure~\ref{fig:good_cases}d shows that our approach is not affected by the background image.

To present a usage example of these extensible templates,
we further implement a timeline renderer 
by extending TS~\cite{timelineStoryteller},
an open-source tool that allows users 
to generate timelines for their data automatically.
It embeds a collection of heuristics to render timelines 
based on the timeline \emph{representation}, \emph{scale}, and \emph{layout} chosen by users.
Currently, TS provides a set of default styles 
(\emph{e.g.}, using \emph{rectangles} as event marks).
We reuse and extend the heuristics in the tool 
and adapt it to our templates,
thus enabling generations of embellished timeline infographics.
We also add some heuristic rules for effectively using marks in templates, 
such as looping through the marks when the number of events exceeds that of the marks.
We present two examples to illustrate the generation process.

\begin{itemize}[leftmargin=*]

\begin{figure}[h]
    \centering 
    \includegraphics[width=\columnwidth]{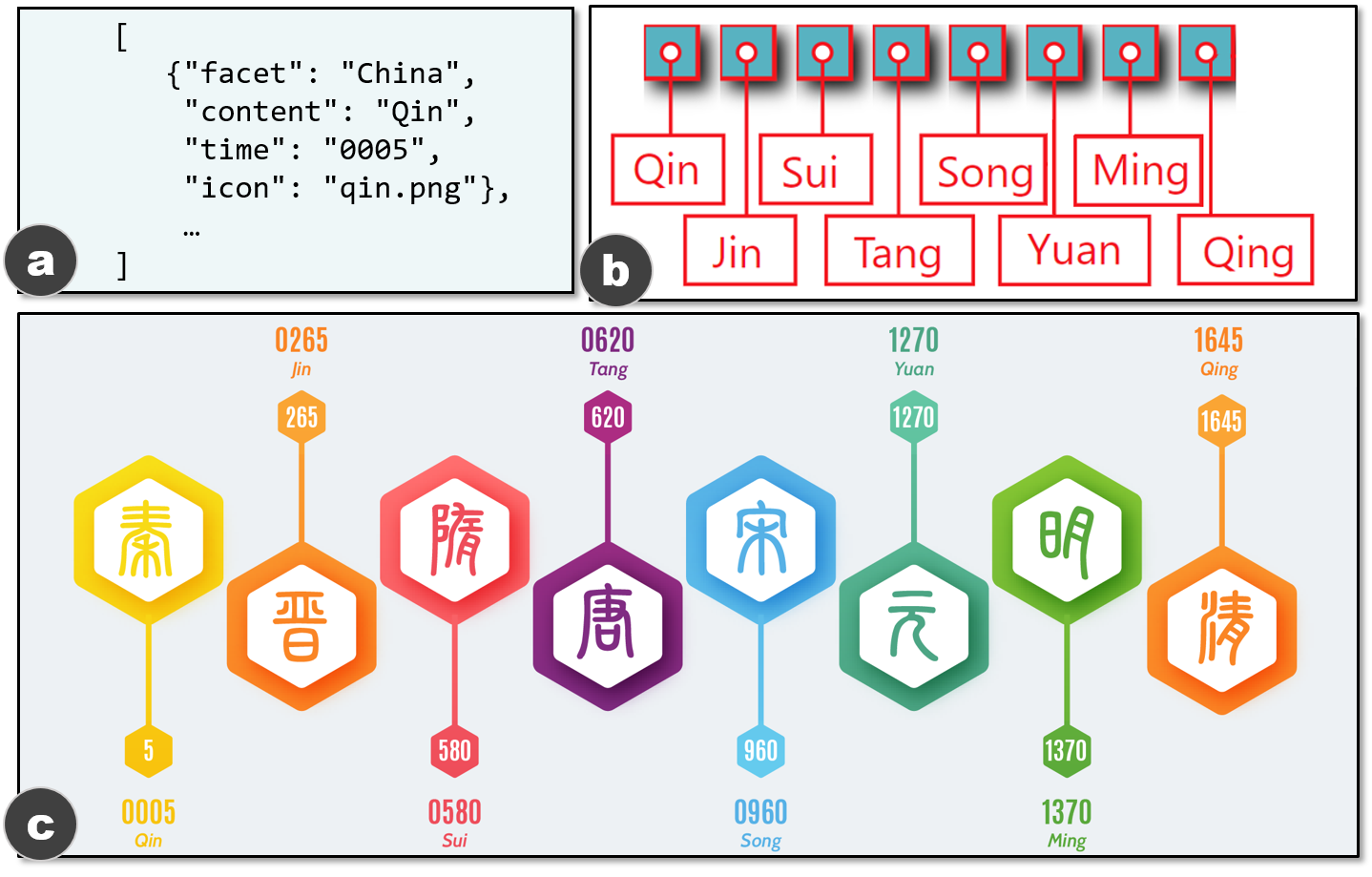}
    \caption{
        A default timeline generated for the data in a) is shown in b).
        The result of applying the template from \autoref{fig:good_cases}a is presented in c).
    }
    \label{fig:generated_cases_a}
\end{figure}

    \item \textbf{Generating timeline reusing graphical elements.}
    In this example, we reuse the graphical elements in \autoref{fig:good_cases}a 
    to generate a new timeline (\autoref{fig:generated_cases_a}c) 
    for the Chinese dynasties data (\autoref{fig:generated_cases_a}a).
    To achieve this,
    we first use TS to render the data 
    with default styles (\autoref{fig:generated_cases_a}b).
    The template is then applied to the default timeline and the underlying data.
    Specifically,
    we first substitute 
    marks in \autoref{fig:generated_cases_a}b (\emph{e.g.}, the rectangles and red circles) 
    with
    the marks in the template (\emph{e.g.}, event and annotation marks),
    which are segmented from the existing timeline.
    The margin and position of each mark are also updated accordingly.
    Then, 
    the event content is rendered at the position corresponding to each event mark
    using the font from the template (\emph{i.e.}, annotation text).
    Next,
    the time of each event is rendered, 
    though it has not been visualized in \autoref{fig:generated_cases_a}b,
    since it exists in the data 
    and the template has slots for the event text.
    Finally, the icon (\emph{i.e.}, the Chinese character) 
    in the data is displayed 
    in the bbox of the annotation icon for each event.
    This whole process is finished programmatically and automatically
    by enumerating and binding the data to the template,
    as we have the semantic role (\emph{i.e.}, the category)
    of elements in the template.

\begin{figure}[ht]
    \vspace{-1mm}
    \centering 
    \includegraphics[width=\columnwidth]{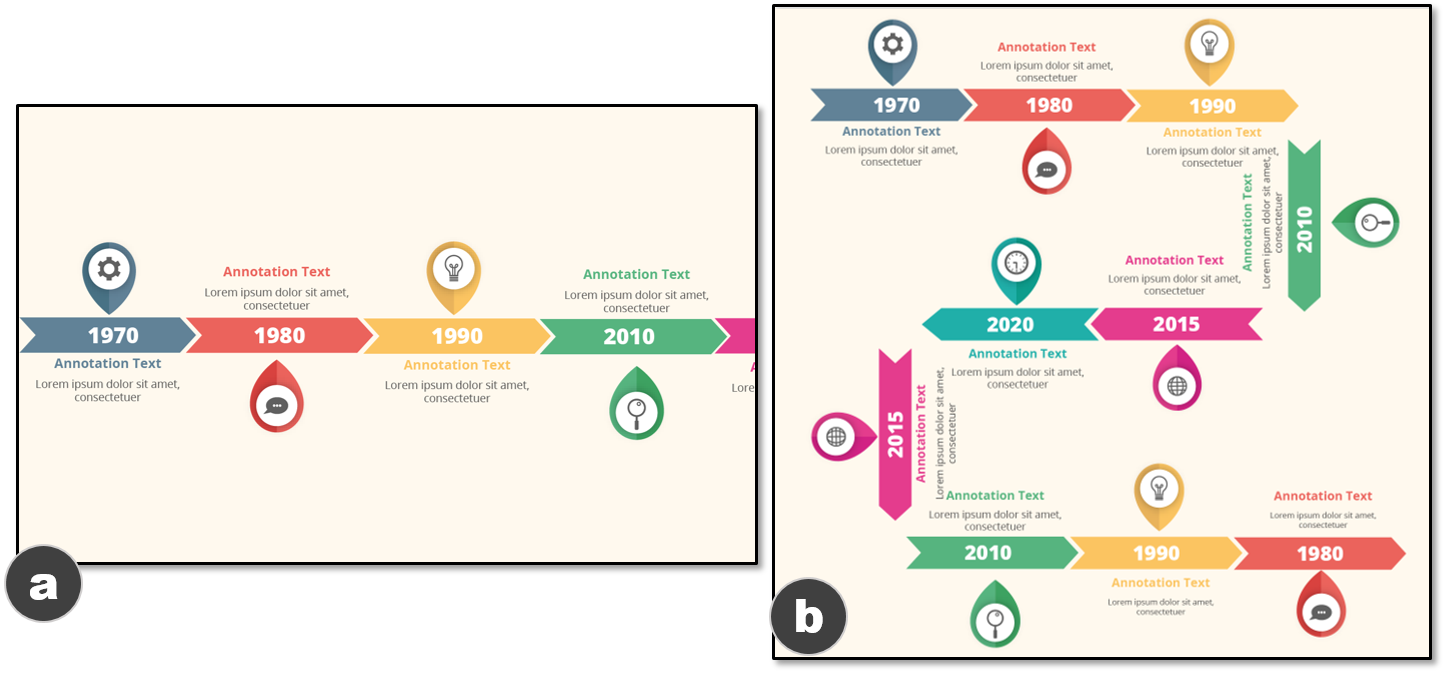}
    \caption{
        a) A timeline generated by using the template from \autoref{fig:good_cases}b. \\
        b) The result of applying the representation from \autoref{fig:good_cases}f to a).
    }
    \label{fig:generated_cases_b}
    \vspace{-1mm}
\end{figure}

    \item \textbf{Generating timeline reusing representations.}
    In this example, we reuse the representation in \autoref{fig:good_cases}f
    to generate a new timeline (\autoref{fig:generated_cases_b}b) for mock-up data.
    To this end,
    we first render a timeline (\emph{i.e.}, \autoref{fig:generated_cases_b}a) 
    using the graphical elements from \autoref{fig:good_cases}b,
    following the same process mentioned above.
    The \emph{arbitrary} representation extracted from \autoref{fig:good_cases}f
    is then reused 
    in \autoref{fig:generated_cases_b}b.
    Specifically,
    we enumerate events in \autoref{fig:generated_cases_b}a
    and place them one by one 
    according to the position (represented by bbox)
    of elements in \autoref{fig:good_cases}f.
    Finally, two events are adjusted to be vertical 
    after the generation process.

\end{itemize}


\section{Discussion}


\cmo{
In this section,
we first share the lessons learned
from our study,
which outline the need for 
\emph{human-ML collaborative authoring tool} and}
\emph{graphical image-driven deep learning}.
Then, we discuss how our work can serve as a basis for future research
and acknowledge the limitations. 

\subsection{Human-ML Collaborative Authoring Tool}
\label{ssec:human_dl}

    Efforts have been made to aid visualization authoring by using ML.
    As a pioneering work, 
    ReVision~\cite{Savva2011} uses SVM and multiple rule-based methods 
    (\emph{i.e.}, computational interpretation)
    to parse the global and local information of bitmap charts. 
    This information is used to help users to redesign problematic charts. 
    Although ReVision successfully decomposes charts, 
    its rules limit its extendibility to tackle more complex visualizations (\emph{e.g.}, infographics). 
    In recent years, 
    given the rapid advances in DL,
    it is gradually possible to use data-driven models
    rather than rule-based methods
    to interpret charts and even infographics.
    In this work, we explore this direction and contribute a unified model to successfully parse timeline infographics.

    Note that our approach aims to use automation to assist, 
    not replace, 
    human designers in the visualization design.
    We realize that it is 
    important to keep the human in
    the design process.
    This is
    not only because it is difficult
    to get a perfect model,
    but also because the design process is creative and subjective, 
    and thus can hardly be fully automatic.
    First, human interactions can steer the model and refine the model's results.
    For example, 
    although we design several automatic post-processing steps 
    to enhance the overall performance of our model,
    the user can further adjust the model (\emph{e.g.}, the confidence level) for the desired output.
    Furthermore,
    the human should be at the center of visualization authoring, 
    while the ML model should assist, rather than replace, the designer.
    For instance, 
    the generated results of our approach can aid users 
    as stepping stones to initialize the visualization, 
    instead of being the final designs.
    However, 
    designing human-ML collaborative authoring tools 
    that go beyond asking designers to refine the model results
    remains underexplored.
    We envision
    how to design authoring tools
    seamlessly integrate imperfect ML models into the design process
    as an important research direction.

\subsection{Graphical Image-Driven Deep Learning}
The DL revolution is driven by tasks on natural images.
However, the specificity of graphical images (\emph{e.g.}, charts and infographics) 
leads to requirements that cannot be easily fulfilled
by models designed for natural images.
We share the lessons learned from our study
and hope to inspire more future work on the fundamental designs of DL models.

\vspace{1mm}
\textbf{Translation invariance \emph{vs.} translation variance}
In some cases,
our model cannot distinguish event marks from annotation marks, when they look identical. 
Although our reconstruction pipeline can fix such incorrect classification in most cases
by using Redundancy Recovery, 
we note that this issue is caused by a key feature of CNNs, namely, \emph{translation invariance}.
Translation invariance~\cite{Goodfellow2016} enables a CNN to recognize an object 
wherever it is displayed in an image.
This feature is important in recognizing natural elements (\emph{e.g.},
a cat should always be classified as ``cat'' wherever it is displayed).
However, it is difficult to handle graphical images, as some graphical elements are translation-invariant while others are translation-variant. 
For instance, in a bar chart image,
the bars in the plot area should always be classified as ``bar mark'',
which requires translation invariance;
by contrast,
text labels' roles are usually determined by their positions 
(\emph{e.g.}, ``y-axis label'' at the left and ``x-axis label'' at the bottom)
and thus requires translation variance.
A possible solution to this problem is to learn and recognize relationships among elements.
Capsule network~\cite{Sabour2017} is a network structure 
that can learn the relationships among elements.
Further investigation is required to adapt it to graphical images.

\textbf{High-level semantics \emph{vs.} low-level semantics} 
Our network can predict the pixel-wise masks of elements,
but their quality is far from perfect.
The problem is rooted in the difference between natural and graphical elements.
In general, natural elements do not have crisp edges.
Thus, most of the models are designed to use low-resolution, semantically-strong features
for improved detection,
while the precision of segmentation is compromised.
By contrast,
graphical elements require precise segmentation because of their crisp edges,
while high-level semantics are still necessary for detection.
This demands a high-resolution, semantically strong features,
which is non-trivial to attain.
One possible future direction is to use various features for various purposes:
low-resolution, semantically strong features for detection;
and high-resolution, semantically weak features for segmentation.

\begin{figure}[ht]
\centering 
    \vspace{-1mm}
    \includegraphics[width=\columnwidth]{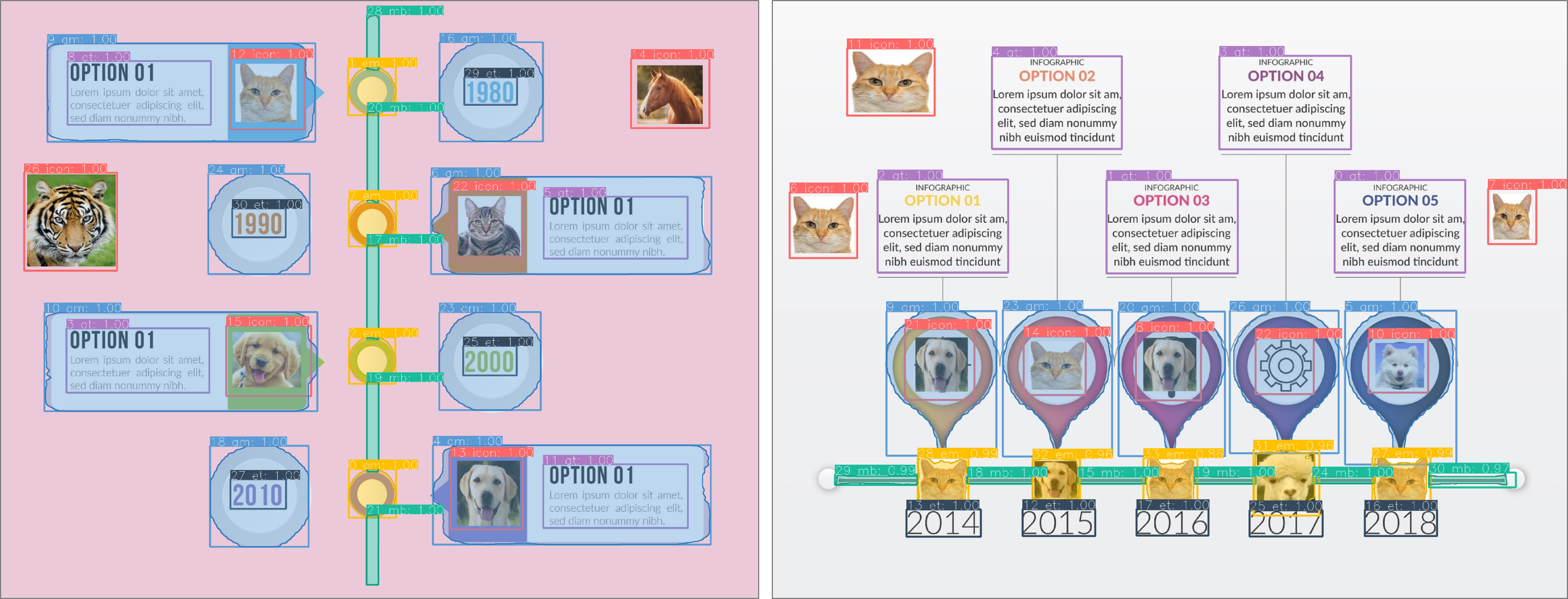}
    \caption{The animals can be correctly identified as timeline elements.
    }
    \label{fig:hybrid}
    \vspace{-2mm}
\end{figure}

\textbf{Single \emph{vs.} hybrid}
Another difference between natural and graphical images
is that graphical images can comprise natural elements and graphical elements.
For example,
in an infographic,
a common practice is to show objects with photos
and annotate them with graphical shapes.
Such kind of hybrid components requires a model 
that considers the characteristics of natural and graphical elements.
However, some of these characteristics may lead to conflict design requirements,
and result in challenges in design models.
Although our datasets do not include natural elements,
we are interested in the performance of our model
on timelines contain graphical and natural elements.
Thus, we randomly substitute some graphical marks
with photos of animals and then feed them to our model.
The results show that our model can still correctly classify the categories of these animals (\autoref{fig:hybrid}).
We regard this performance as a benefit of the pre-trained network.
Future research is needed to further understand the generality of these cases.

\subsection{Future Work}
We propose 
an end-to-end approach to automatically extract extensible templates from infographic timelines. 
This work is only a starting point towards
automated infographic design.
Here, we discuss some promising opportunities 
that can facilitate the design process of infographics.

\textbf{From timeline to the others.} 
Although our approach focuses on infographic timelines,
it can be generalized to other cases.
First, 
our model can be extended to more than 10 types of timelines
once a larger and more diverse dataset is available.
The ability of our model to parse infographics mainly 
depends on the training dataset rather than the rules. 
Furthermore, the two tasks, namely, 
parsing the global and local information of a visualization, 
are not specific to timeline infographics
but applicable to other types of visualizations. 
For example,
previous research~\cite{Savva2011, Poco2017} on charts decomposition 
also contains similar steps to parse the global and local information. 
Given that our model is data-driven, 
it is possible to train the model using 
other types of visualizations, 
thereby extending the model to a broader scenario. 
Finally, besides facilitating visualization authoring, 
the templates can be used to make static infographics interactive, 
\emph{e.g.}, using the parsed information to support selection and filtering of elements.
The parsed information can also be used for other applications,
such as indexing of infographics,
retargeting of visual styles, 
and infographics content analysis.

To sum up, the generalizability of our model is limited by the training data.
Generally speaking, DNN models, 
no matter supervised or unsupervised, 
require a large amount of training data to achieve high prediction accuracy. 
For the scenarios where annotated datasets are unavailable, 
our model has limited applicability.
We plan to extend the application scenarios of our approach in the future.


\textbf{From hybrid to purely learning-based.}
Although our approach utilizes a heuristic pipeline,
it can be improved by substituting the pipeline with a DL model.
Our work shows that after extracting the features of an input image,
we can decode different image information
(\emph{e.g.,} global and local information) 
by using multi-functional heads.
Given that an extensible template can be represented by a structural document (similar to Vega specifications),
a potential improvement is to use a recurrent neural network (RNN)
to decode the feature maps
and directly output extensible templates.
Recently, Dibia and Demiralp~\cite{Dibia2018} showed
the possibility of translating a JSON-encoded data into Vega-lite specification by using a RNN.
Research in \cmo{CV} field
also presents models that take images as inputs and return textual data as outputs.
The related work suggests the potential to extend our model to an end-to-end model,
which takes infographic images as inputs and directly outputs templates.
We consider this area as an important future direction.

\vspace{1mm}
\textbf{From template-based to freeform.}
Lastly,
our model shows the ability to learn and understand the content of infographic images.
This characteristic indicates several potential directions
to facilitate the design process.
For example,
we can use a trained model to interpret a sketch 
or materials (\emph{e.g.}, data, icons, and textual description) from users 
and recommend infographic templates.
Another step in this direction
is to design mixed initiative authoring systems,
including automatically completing or generating designs on the basis of users' input.

\subsection{Limitations}
We acknowledge the limitations of our study.
First,
our datasets are rather limited while the generalizability of our model mainly depends on the data.
However, collecting high-quality infographic datasets is not an easy task
considering the manual labeling efforts.
We plan to open source~\footnote{https://chenzhutian.org/auto-infog-timeline} our datasets and labeling tools for the community
and collect larger-scale infographic datasets in the future.
Second,
given our work is not aimed at high metric values,
we did not optimize our model with bells and whistles,
including multi-scale train/test, OHEM~\cite{Shrivastava2016},
and other techniques.
Outside the scope of this work, we expect that such improvement skills are applicable to our model.
Third, although our approach can automatically extract templates from infographic timelines,
its performance can be further improved by involving users' refinements.
For example, we can integrate our approach to infographic authoring tools
and thus allow users to interactively refine extracted results.
A separate limitation is that the learning process is a 'blackbox',
which calls for investigations on the learning process.
Finally,
when applying our approach for some purposes (\emph{e.g.}, reusing graphical elements or overall layouts), 
we suggest that users should note the copyright issue, 
which is complicated and depends on the law varying among countries,
the purpose of usage (\emph{e.g.}, commercial \emph{vs.} noncommercial), 
the degrees of the redesign, \emph{etc.}

\section{Conclusion}

We contribute an automated approach to extract extensible templates from bitmap infographic timelines.
A multi-task DNN is presented to understand and deconstruct bitmap timeline infographics, 
by classifying the types and orientations of timelines
and detecting and segmenting elements on timelines; 
from these results,
a heuristic pipeline 
is used to reconstruct extensible templates.
The extensible templates can be used to automatically
generate timeline infographics
with updated data.
The quantitative experiments
and example results confirm the effectiveness and usefulness of our approach.
We share lessons learned from our study
which make us notice
the needs of \emph{graphical image-driven deep learning}.
We also discuss how our work
can be extended towards automated infographics design
in future researches.

\acknowledgments{
The authors wish to thank Weiwei Cui for valuable feedback on this project,
Sikai Cai, Zicheng Xu, Kun Xie for assistance in labeling the dataset,
as well as the anonymous reviewers for their valuable comments.
This work is partially supported by a grant from MSRA (code: MRA19EG02).
}

\bibliographystyle{abbrv}
\newpage
\bibliography{template}
\end{document}